\documentclass[aps,prb,twocolumn,superscriptaddress,amsmath,amssymb]{revtex4-2}

\usepackage{graphicx}
\usepackage{lipsum}

\usepackage{hyperref}  
\usepackage[all]{hypcap} 

\usepackage{color}

\newcommand{\chem}[1]{\ensuremath{\mathrm{#1}}} 
\newcommand{\co}{Co_3O_2BO_3}
\newcommand{\sn}{Co_5Sn(O_2BO_3)_2}

\newcommand{\three}{Co^{3+}}

\usepackage{float}

\begin{document}

\title{Structural and spectroscopic investigation of the charge-ordered, short-range ordered, and disordered phases of the \ensuremath{\boldsymbol{\mathrm{Co_3O_2BO_3}}} ludwigite}

\author{C. W. Galdino}
\affiliation{``Gleb Wataghin'' Institute of Physics, University of Campinas (UNICAMP), Campinas, S\~ao Paulo, 13083-859, Brazil}

\author{D. C. Freitas}
\affiliation{Instituto de F\'isica, Universidade Federal Fluminense, Campus da Praia Vermelha, Niter\'oi, RJ, 24210-346, Brazil}

\author{C. P. C. Medrano}
\affiliation{Instituto de F\'isica, Universidade Federal Fluminense, Campus da Praia Vermelha, Niter\'oi, RJ, 24210-346, Brazil}

\author{D. R. Sanchez}
\affiliation{Instituto de F\'isica, Universidade Federal Fluminense, Campus da Praia Vermelha, Niter\'oi, RJ, 24210-346, Brazil}

\author{R. Tartaglia}
\affiliation{``Gleb Wataghin'' Institute of Physics, University of Campinas (UNICAMP), Campinas, S\~ao Paulo, 13083-859, Brazil}

\author{L. P. Rabello}
\affiliation{Instituto de F\'isica, Universidade Federal Fluminense, Campus da Praia Vermelha, Niter\'oi, RJ, 24210-346, Brazil}

\author{A. A. Mendonça}
\affiliation{Instituto de F\'isica, Universidade Federal do Rio de Janeiro, Caixa Postal 68528, Rio de Janeiro, RJ, 21941-972, Brazil}

\author{L. Ghivelder}
\affiliation{Instituto de F\'isica, Universidade Federal do Rio de Janeiro, Caixa Postal 68528, Rio de Janeiro, RJ, 21941-972, Brazil}

\author{M. A. Continentino}
\affiliation{Centro Brasileiro de Pesquisas F\'isicas, Rua Dr. Xavier Sigaud, 150 - Urca, Rio de Janeiro, RJ, 22290-180, Brazil}

\author{M. J. M. Zapata}
\affiliation{Departamento de F\'isica, Universidade Federal de Minas Gerais, Belo Horizonte, MG, 31270-901, Brazil}

\author{C. B. Pinheiro}
\affiliation{Departamento de F\'isica, Universidade Federal de Minas Gerais, Belo Horizonte, MG, 31270-901, Brazil}

\author{G. M. Azevedo}
\affiliation{Instituto de F\'isica, Universidade Federal do Rio Grande do Sul (UFRGS), Porto Alegre, RS, 90040-060, Brazil}

\author{J. A. Rodr\'iguez-Velamaz\'an}
\affiliation{Institut Laue Langevin, 38042 Grenoble, France}

\author{G. Garbarino}
\affiliation{European Synchrotron Radiation Facility, 38043 Grenoble, France}

\author{M. N\'u\~nez-Regueiro}
\affiliation{Institut N\'eel/CNRS-UJF, 38042 Grenoble, France}

\author{E. Granado}
\affiliation{``Gleb Wataghin'' Institute of Physics, University of Campinas (UNICAMP), Campinas, S\~ao Paulo, 13083-859, Brazil}

\date{\today}

\begin{abstract}
Charge-ordering is prone to occur in crystalline materials with mixed-valence ions. It is presumably accompanied by a structural phase transition, with possible exceptions in compounds that already present more than one inequivalent site for the mixed-valence ions in the charge-disordered phase. In this work, we investigate the representative case of the homometallic Co ludwigite Co$^{2+}_2$Co$^{3+}$O$_2$BO$_3$ ($Pbam$ space group) with four distinct Co crystallographic sites [$M1$--$M4$] surrounded by oxygen octahedra. The mixed-valent character of the Co ions up to at least $T=873$~K is verified through x-ray absorption near-edge structure (XANES) experiments. Single crystal x-ray diffraction (XRD) and neutron powder diffraction (NPD) confirm that the Co ions at the $M4$ site are much smaller than the others at low temperatures, consistent with a Co$^{3+}$ oxidation state at $M4$ and Co$^{2+}$ at the remaining sites. The size difference between the Co ions in the $M4$ and $M2$ sites is continuously reduced upon warming above $\approx 370$~K, indicating a gradual charge redistribution within the $M4$--$M2$--$M4$ (424) ladder in the average structure. Minor structural anomalies with no space group modification are observed near 475 and 495~K, where sharp phase transitions were previously revealed by calorimetry and electrical resistivity data. An increasing structural disorder, beyond a conventional thermal effect, is noted above $\approx 370$~K, manifested by an anomalous increment of XRD Debye-Waller factors and broadened vibrational modes observed by Raman scattering. The local Co--O distance distribution, revealed by Co $K$-edge Extended X-Ray Absorption Fine Structure (EXAFS) data and analyzed with an evolutionary algorithm method, is similar to that inferred from the XRD crystal structure below $\approx 370$~K. At higher temperatures, the local Co--O distance distribution remains similar to that found at low temperatures, at variance with the average crystal structure obtained with XRD. We conclude that the oxidation states Co$^{2+}$ and Co$^{3+}$ are instantaneously well defined in a local atomic level at all temperatures, however the thermal energy promotes local defects in the charge-ordered configuration of the 424 ladders upon warming. These defects coalesce into a phase-segregated state within a narrow temperature interval ($475< T < 495$~K). Finally, a transition at $\approx 500$~K revealed by differential scanning calorimetry (DSC) in the iron ludwigite Fe$_3$O$_2$BO$_3$ is discussed. 
\end{abstract}

\maketitle

\section{Introduction}

Charge-ordering, namely the long-range arrangement of different oxidation states of the same chemical species in a given crystalline material, has been widely studied since it was first proposed to explain the so-called Verwey transition at $T_c=120$~K in magnetite (Fe$_3$O$_4$) \cite{verwey1939electronic, Iizumi1982}. Since then, charge-ordering have been identified as the driving mechanism of structural phase transitions in a number of transition-metal oxides \cite{coey2004,Attfield2006}. Indeed, if the metal ion $M$ occupy a single crystallographic site in the high-temperature parent structure, a symmetry reduction is necessary to accommodate the low-temperature charge-ordered state, and a combined charge-order/structural phase transition occurs at a given $T_c$ \cite{coey2004,Garcia2004}. On the other hand, materials showing inequivalent $M$ crystallographic sites in the parent structure are candidates for developing charge-ordered and disordered states under the same space group, providing opportunities to investigate solely the charge-ordering phenomenon without the accompanying structural phase transition. In addition, since no symmetry breaking is associated with the charge-ordering in these cases, a gradual charge-order melting crossover rather than a phase transition may occur, paving the way for intermediate configurations that would be otherwise concealed by the phase transition. Therefore, a closer look at this class of materials may reveal so-far unexplored physics related to the charge-ordering process and its intermediate steps.

\begin{figure}
	\centering
	\includegraphics{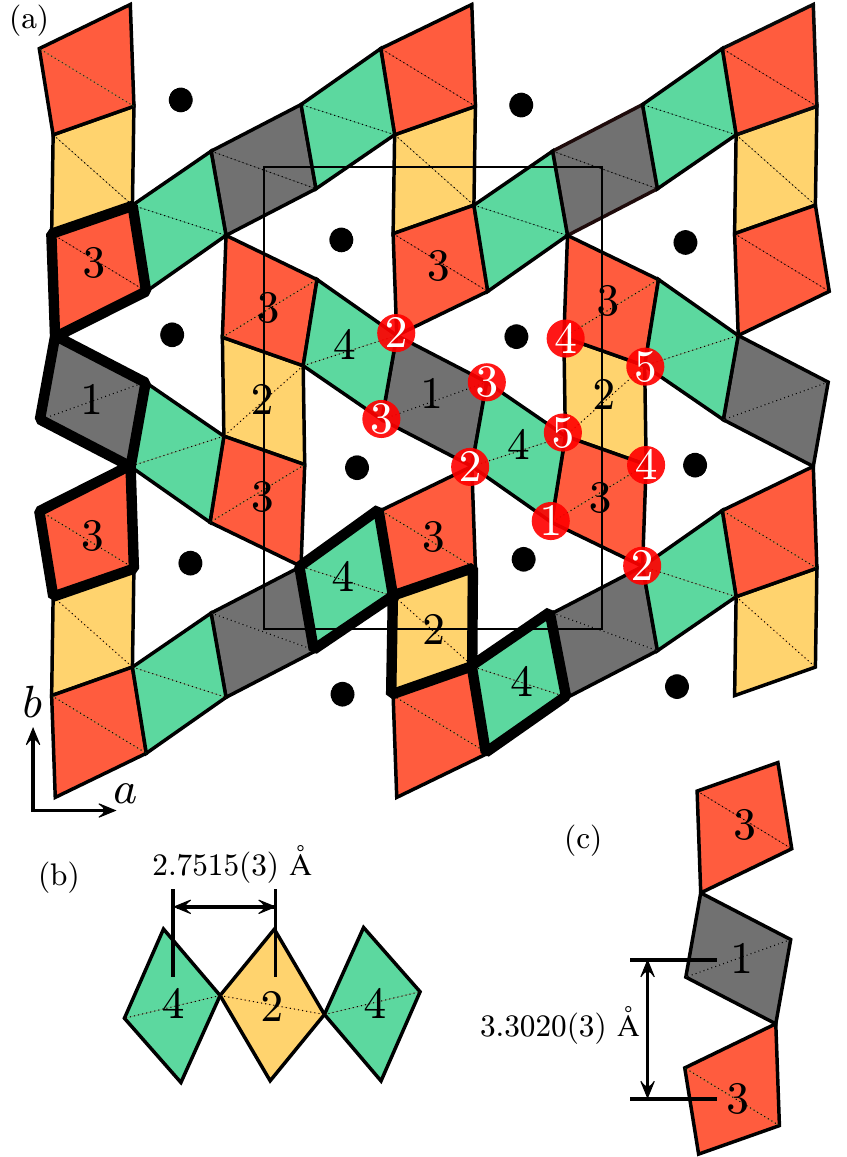}
	\caption{(a) Polyhedral representation of the ludwigite structure (space group $Pbam$) projected in the $ab$ plane. This arrangement is simply repeated along the {\bf c} direction. Metal sites $M$ are numbered from 1 to 4. The five oxygen sites are indicated by red circles and black dots represent the boron ions. The solid rectangle represents the unit cell. Thick lines indicates two crystal subunits, namely the 424 and 313 three-legged ladders that run along {\bf c}, which are also displayed in more detail in (b) and (c), respectively, with the corresponding $M$--$M$ average bond length obtained with x-ray diffraction data at 300~K.} 
	\label{fig_structure1}
\end{figure}

Ludwigites with chemical formula $M^{2+}_2(M')^{3+}$\chem{O_2BO_3} are oxyborates with oxygen octahedra surrounding the transition-metal ions $M$ and $M'$ occupying four different crystallographic sites $M1$--$M4$ (see Fig.~\ref{fig_structure1}). So far, the only two known homometallic ($M=M'$) ludwigites with mixed-valence $M^{2+}$ and $M^{3+}$ ions are \chem{\co} and Fe$_3$O$_2$BO$_3$. The structure and magnetism of these materials are often rationalized in terms of $M4$--$M2$--$M4$ (424) triads (also called three-legged ladders) with the shortest $M$--$M$ distances, and $M3$--$M1$--$M3$ (313) triads or ladders with the longest $M$--$M$ distances (see Fig.~\ref{fig_structure1}). Particularly, the $M$=Fe ludwigite shows a combined charge-ordering and structural phase transition at 283~K involving a zigzag distortion of the 424 ladders below this temperature~\cite{douvalis2002mossbauer, guimaraes1999cation, larrea2001charge, mir2001structural, fernandes2005transport}. The $M$=Co ludwigite, on the other hand, does not show a similar structural phase transition, and all Co$^{3+}$ ions appear to occupy only the $M4$ site~\cite{freitas2008structure,freitas2016magnetism,kazak2021spin}. Additionally, there is a possible spin-state crossover of the \chem{\three} ions, which are presumably in a low-spin configuration in the magnetically ordered phase ($T \leq 43$ K) and in a high-spin configuration at higher temperatures ($T \gtrsim 200$ K) where the Curie-Weiss paramagnetic behavior is established~\cite{freitas2016magnetism, galdino2019magnetic, matos2020low}. Indeed, the combination of low dimensional units and mixed valence states leads to a number interesting physical phenomena in ludwigites \cite{continentino2001magnetic, sofronova2017ludwigites}, such as dimerized states~\cite{mir2001structural}, structural and charge-ordering transitions~\cite{mir2001structural, sanchez2004magnetism, fernandes2005transport, bordet2009magnetic, galdino2019magnetic}, spin-glass states~\cite{kazak2009low, freitas2009partial, kumar2017reentrant, ivanova2012spin, knyazev2012crystal, heringer2020spin, kulbakov2021destruction, freitas2010structural, medrano2021magnetic}, and multiferroicity \cite{damay2021high}. Also, these oxyborates have been tested as high-performance electrodes for lithium-ion and sodium-ion batteries~\cite{sottmann2018playing, zhou2019carbon, ping2019iron, pralong2017electrochemical}, low frequency oscillators~\cite{dos2016current, dos2017non}, and as oxygen evolution reaction (OER) electrocatalyst~\cite{kundu2018mixed}.

A previous study showed that \chem{\co} presents anomalies in the heat flow and resistivity data at 475 and 495~K, revealing sharp phase transitions that do not appear to be structural in nature even though some lattice anomalies were perceived \cite{galdino2019magnetic}. The $a$ lattice parameter shows an anomalous contraction upon warming through a broad temperature interval between $\approx 370$ and 600~K, whereas the $c$ lattice parameter shows subtle anomalies at 475 and 495~K superposed with a broad anomalous behavior between $\approx 370$ and 500~K. These anomalies were ascribed to a charge-order melting process where trivalent ions, once located at Co site 4, shares their extra positive charge with adjacent divalent Co ions \cite{galdino2019magnetic}. The anomalous behavior of the lattice parameters and discontinuities in heat flow and resistivity measurements indicate interesting physics and justify a more detailed structural investigation. Here, we investigate the crystal, local atomic and local electronic structures as well as the vibrational properties of \chem{\co} by means of a combination of crystallographic and spectroscopic techniques, namely, single-crystal x-ray diffraction (XRD), neutron powder diffraction (NPD), extended x-ray absorption fine structure (EXAFS), x-ray absorption near-edge structure (XANES) and Raman scattering, over a wide temperature interval (from 6 to 873~K). This combination of techniques provides insight into the gradual charge-order/disorder process, leading to a simple picture to explain the previously observed sharp high-temperature phase transitions in this compound involving an intermediate state. This picture also provides insight into the physics of the other homometallic ludwigite Fe$_3$O$_2$BO$_3$ at high temperatures.

\section{Experimental and computational details}
\label{sec_experiment}

Pure Co ludwigite~\chem{\co} and Sn doped~\chem{\sn} single crystals were synthesized according to Freitas et al.~\cite{freitas2008structure} and Medrano et al.~\cite{medrano2015nonmagnetic}, respectively. The single crystals were acicular (needle-shaped) with length of roughly 0.7~mm and the cross-section may be approximated by a square with sides of length 50~$\mu$m.

X-ray diffraction data below room temperature were collected with a Bruker D8-Venture diffractometer with Mo K$\alpha$ radiation ($\lambda$~=~0.71073~\AA) and a nitrogen gas Oxford Cryostream Cooler cryostat. A single crystal was mounted on a Kappa goniometer and the data collection was performed with APEX3 \cite{bruker2016apex3}. Above room temperature, x-ray diffraction data were collected on an Oxford-Diffraction GEMINI-Ultra diffractometer also using Mo K$\alpha$ radiation and an Enraf-Nonius furnace. CRYSALIS PRO software \cite{crysalis2015rigaku} was used for data integration, scaling of the reflections and analytical absorption corrections. Space group identification was done with XPREP \cite{sheldrick2015crystal}, and structure solution was carried out by direct methods using SUPERFLIP \cite{palatinus2007superflip}. For both data sets, multi-scan correction using equivalent reflections was applied, where the full-matrix least-squares refinements based on F$^2$ with anisotropic thermal parameters were carried out using WINGX \cite{farrugia2012wingx} and SHELXL packages \cite{sheldrick2015crystal}. NPD measurements were carried out in the D20 instrument of the Institut Laue-Langevin~(ILL) in Grenoble with $\lambda$~=~1.36~\AA. Ground single crystals were kept in a cylindrical vanadium can and put inside a furnace. Experimental data were analyzed using the FULLPROF Suite programs \cite{rodriguez1993fullprof}.

Heat flow measurements were performed using a differential scanning calorimeter, model Q2000 from TA Instruments, following a heat/cool/heat procedure under a cooling/heating rate of 10~K/min between 250 and 550~K. A linear baseline attributed to the instrumental contribution was subtracted from the raw experimental data.

Raman spectra of a \chem{\co} single crystal were collected using a Jobin-Yvon triple 1800~mm$^{-1}$ grating spectrometer with a $L$N2-cooled charge coupled device and a closed-cycle He cryostat operating up to $T=500$~K. The excitation source was a 532~nm Cobolt diode laser focused in a single crystal in a nearly backscattering geometry with spot size of $\approx 50 \mu$m working at 20~mW. The incident light was polarized perpendicular to the crystal's $c$ direction and analyzed with the same polarization. Finally, fitting of the experimental data was performed using a python-script-based non-linear least square method provided by the Scipy package~\cite{jones2001scipy}, where spectral features were fitted using Lorentzian profiles.

X-Ray absorption spectra at the Co $K$-edge of \chem{\co}, \chem{\sn} and standard Co compounds (\chem{CoO} and \chem{LaCoO_3}) were measured at LNLS at the XAFS2 beamline for various temperatures using the transmission mode. A bunch of single crystals was ground until a fine powder was obtained, which was sieved with a $10$~$\mu$m pore size nylon filter and attached to a circular membrane made of polyvinylidene fluoride through a filtration process. The samples were placed on the cold finger of a closed-cycle He cryostat. For high-temperature measurements, the sieved samples were diluted in boron nitride and molded into a circular pellet required for the homemade furnace. The EXAFS signal of the pelletized samples were slightly damped due to x-ray leakage~\cite{goulon1982experimental} and a correction factor of 1.23, obtained by comparing room-temperature data from pelletized samples and membrane-attached samples, was applied to the high-temperature data. The EXAFS signal before and after the correction is shown in Fig.~\ref{sup_exafs} in the supplementary material (SM). Data normalization and background removal was done through a standard procedure using \textsc{ATHENA}~\cite{ravel2005athena} and energy calibration and alignment was ensured by simultaneous measurements of metallic cobalt. The simulation of EXAFS data was performed by Reverse Monte Carlo calculations attached to a genetic algorithm implemented by EvAX code \cite{timoshenko2014exafs}. The simulated spectrum was adjusted in the R-space from 1 to 4.5~cm$^{-1}$. The initial model structure was formed by 24 unit cells (or 864 atoms) generated from previous diffraction data \cite{cai2014subsolidus},
with lattice parameters for each temperature taken from Refs.~\cite{freitas2008structure, galdino2019magnetic}.

\section{Results and analysis}

\subsection{Single crystal x-ray diffraction}

\begin{figure}
	\centering
	\includegraphics{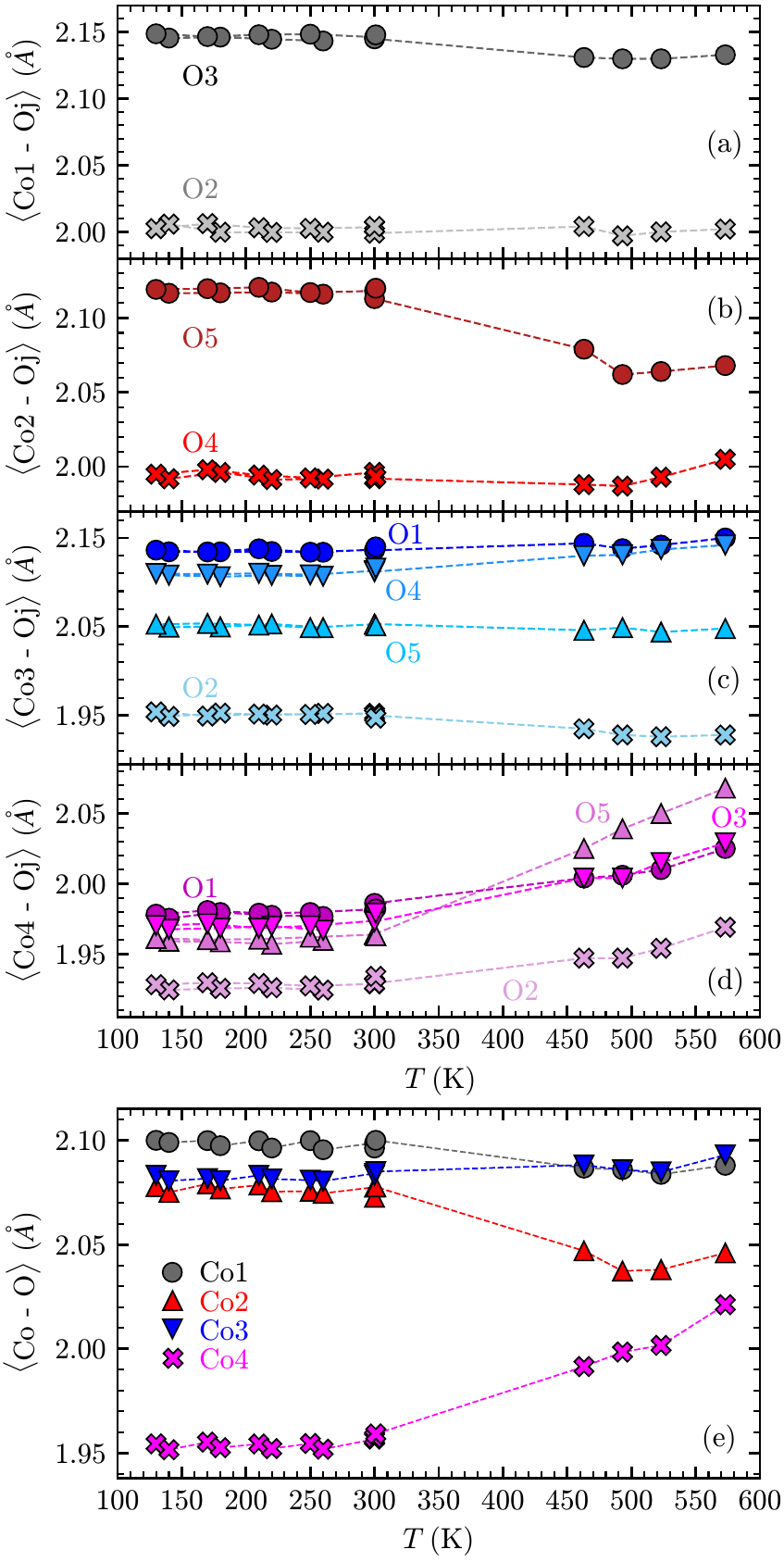}
	\caption{(a--d) Co$i$--O$j$ bond lengths obtained with single crystal x-ray diffraction data. (e) Mean $<$Co$i$--O$>$ bond length for each Co$i$ site averaged over the neighboring oxygen atoms. Statistical errors are smaller than the symbol size
	}
	\label{fig_xrd}
\end{figure}

\begin{figure}
	\centering
	\includegraphics{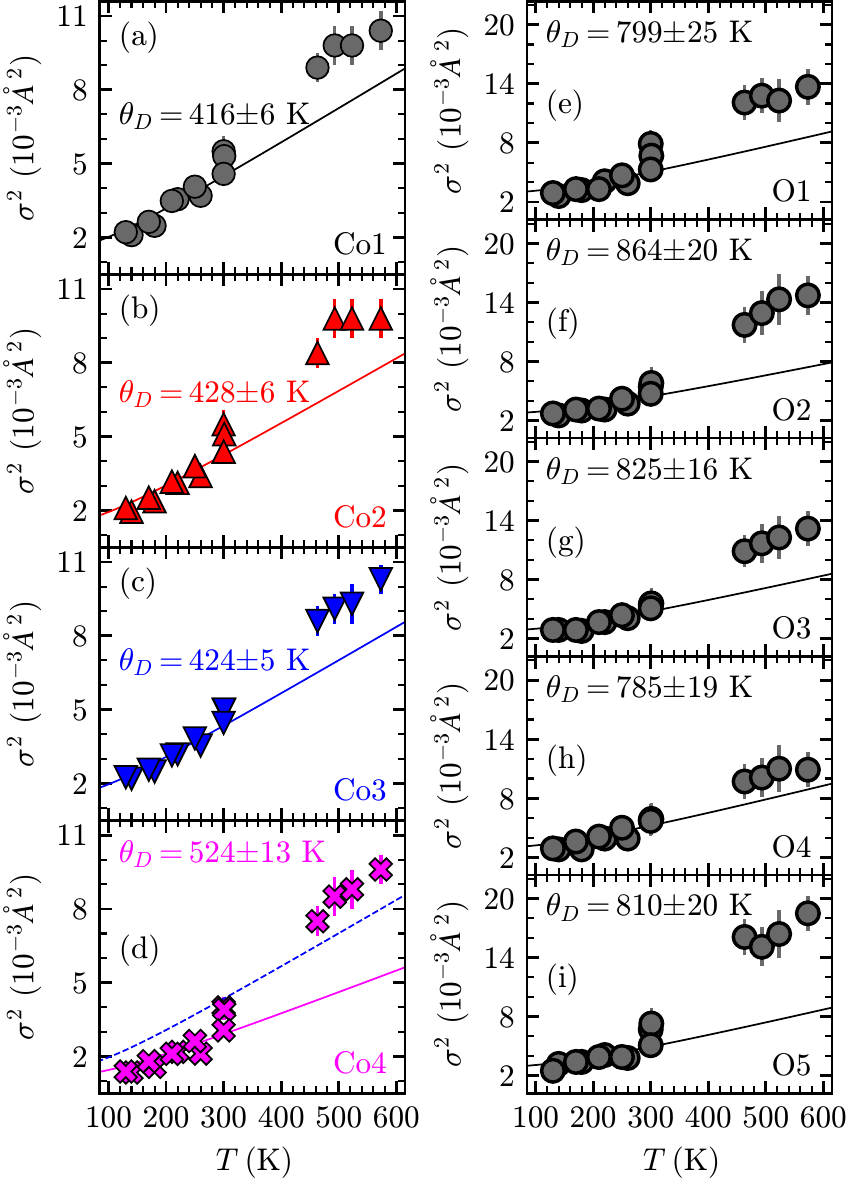}
	\caption{Debye–Waller factors $\sigma^2$ for (a--d) Co1--Co4 and (e--i) O1-O5 sites obtained with single crystal x-ray diffraction data. Solid lines show theoretical curves according to the Debye model (see text) where the value for $\theta_D$ for each site is indicated. The dashed line in (d) is the Debye model fit for site Co3, shown for comparison purposes. 
	}
	\label{fig_debye}
\end{figure}

The crystal structure of \chem{\co} is first investigated with single crystal x-ray diffraction at selected temperatures from 130~K to 573~K. The observed extinction conditions for the $hkl$ reflections are consistent with either the centrosymmetric $Pbam$ or the non-centrosymmetric $Pba2$ space groups over the entire investigated temperature interval. In the $Pbam$ ludwigite structure, all atoms occupy special positions with fixed $z=0$ or $z=1/2$ (Wyckoff positions 2c, 2b, 4g, 4h), whereas this condition is released for the $Pba2$ subgroup. The structure of \chem{\co} could be refined under both space groups, with very similar fitting residuals at all investigated temperatures. We therefore proceed our analysis using the more symmetric $Pbam$ space group, consistent with previous work on ludwigites in general. Finally, Reciprocal space reconstructed layers at selected temperatures are given in the SM and indicate a reflection splitting at 523~K characteristic of crystal twining at that temperature (see Fig.~\ref{sup_twinning} in the SM). Table \ref{sup_table} in the SM shows the refinement data at selected temperatures and the complete crystallographic data for \chem{\co} can be accessed through Cambridge Crystallographic Data Center (CCDC) deposition numbers 2094101-2094115.

Figures~\ref{fig_xrd}(a--d) show the temperature dependence of the Co1--O$j$, Co2--O$j$, Co3--O$j$ and Co4--O$j$ distances, respectively, with $j=1,\ldots,5$ (see also Fig.~\ref{fig_structure1}). Figure~\ref{fig_xrd}(e) shows the mean $<$Co$i$--O$>$ distance for each Co$i$ site ($i=1, \ldots, 4$), which remain nearly temperature-independent from the base temperature (125~K) up to $\approx$~370~K. In this temperature range, $<$Co4--O$>$ $\approx 1.95$~\AA\ is much smaller than the other $<$Co$i$--O$>$ distances (2.08--2.10~\AA). In addition, $<$Co1--O$> = 2.10$~\AA\ is slightly larger than $<$Co2--O$>$ $\approx$ $<$Co3--O$>$ $\approx$ 2.08 \AA. Above 370~K up to 500~K, $<$Co4--O$>$ is substantially incremented, accompanied by an important reduction of $<$Co2--O$>$ by approximately $0.03$~\AA\ and a more modest reduction of $<$Co1--O$>$ by roughly $0.01$~\AA. Figure~\ref{fig_xrd}(b) and Fig.~\ref{fig_xrd}(d) shows that the effect is most directly associated with shortenings of the Co2--O5 and Co1--O3 bonds and elongations of all Co4--O$j$ bonds, where the largest contribution comes from Co4--O5 bond upon warming.


Figures \ref{fig_debye}(a--d) show the temperature dependence of the isotropic Debye-Waller (DW) parameters $\sigma^2$ of the four Co ions, whereas Figs. \ref{fig_debye}(e--i) display $\sigma^2$ for the five oxygen ions of the ludwigite structure. Co1--Co3 ions show similar values of $\sigma^2$ with similar temperature dependencies. The solid lines in Figs.~\ref{fig_debye}(a--d) correspond to fits according to the Debye model~\cite{wood2002thermal},
\begin{equation}
\sigma^2(T) = \frac{145.55 T }{M \theta_D^2} + \varphi\left( \frac{\theta_D}{T}\right)+ \frac{36.39}{M \theta_D}
\end{equation}
where $M$ is the mass of the vibrating ion in atomic mass units and $\varphi$ is given by
\begin{equation}
\varphi\left( \frac{\theta_D}{T}\right) = \frac{T}{\theta_D}\int_0^{\theta_D/T} \left[ \frac{x}{\exp(x) -1}\right]dx
\end{equation}
The fits were performed using only data at $T \leq 300$~K and the modeled curves were then extrapolated to higher temperatures. The Debye temperatures obtained from these fits are $\theta_D = $ $416\pm6$, $428\pm6$~K, $424\pm5$~K, and $\theta_D = 524\pm13$~K for Co1--Co4, respectively. Errors are statistical only and represent one standard deviation. Co4 presents significantly lower $\sigma^2$ and correspondingly higher $\theta_D$ compared to Co1--Co3 at low temperatures. For comparison, the Debye model of site Co3 is also displayed in Fig.~\ref{fig_debye}(d) as a dashed curve. In the high-temperature limit, the $\sigma^2$ values of all Co ions fall above their respective extrapolated DW model, where the $\sigma^2$ values of Co4 shows the largest increment on warming, reaching similar values of the other Co sites. As for the oxygen ions [Fig.~\ref{fig_debye}(e--i)], all sites show similar values of $\sigma^2$ below room temperature with $\theta_D$ ranging from 785~K to 864~K [see Fig.~\ref{fig_debye}(e--i) for exact values]. At higher temperatures, the $\sigma^2$ values of all O ions also fall above the extrapolated DW model. This effect is particularly large for site O5 that connects Co2 and Co4. Overall, it is evident that an enhanced atomic disorder occurs in \chem{\co} at temperatures above room temperature.


\subsection{Neutron powder diffraction}

\begin{figure}
	\centering
	\includegraphics{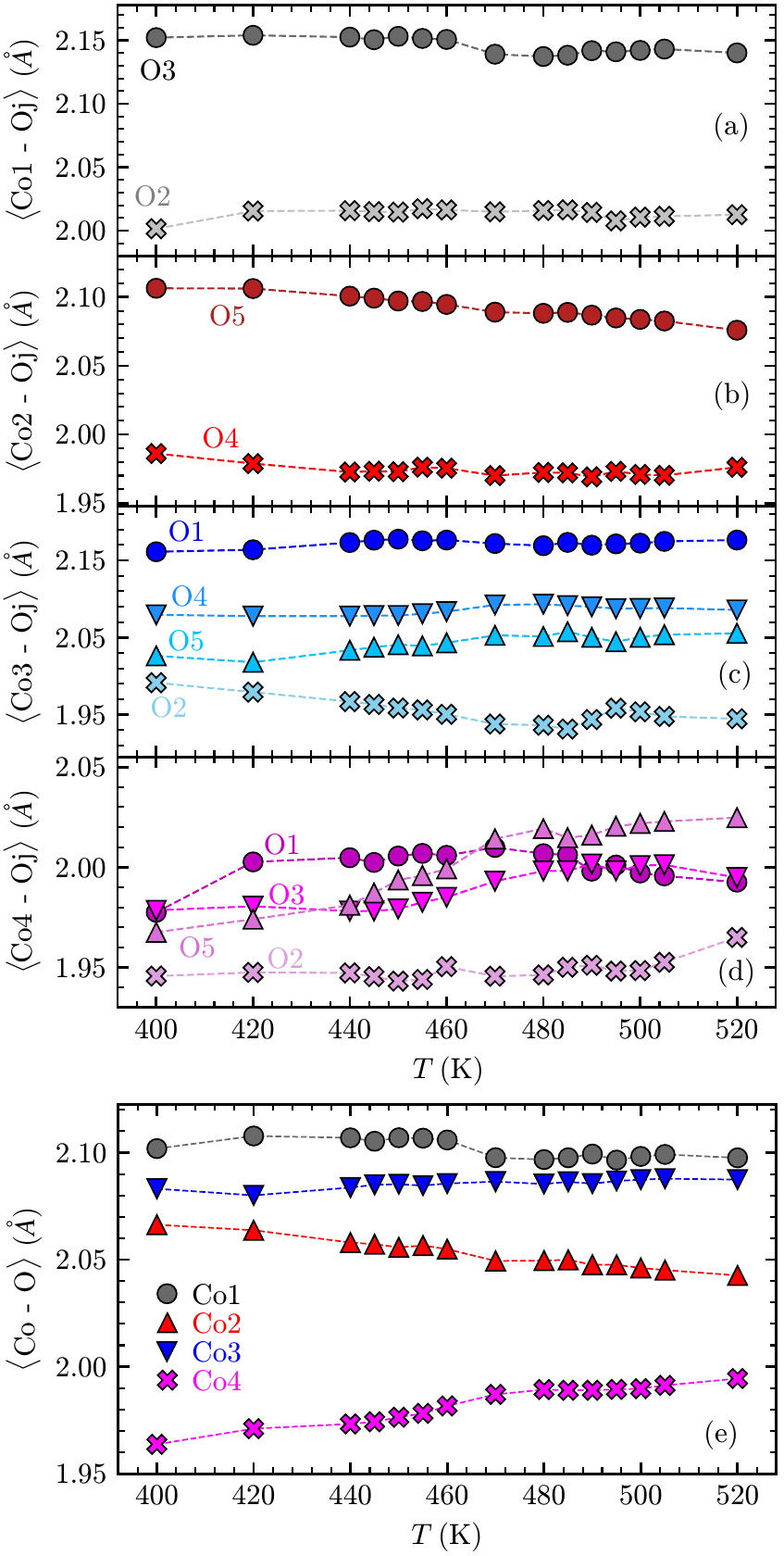}
	\caption{Similar to Fig.~\ref{fig_xrd}, using Rietveld refinement results for neutron powder diffraction data.}
	\label{fig_neutrons}
\end{figure}

Neutron powder diffraction is employed to complement our investigation of the crystal structure of \chem{\co} in a narrower temperature interval ($400<T<520$~K). 
Figures~\ref{fig_neutrons}(a--d) show the temperature dependencies of the Co1--O$j$, Co2--O$j$, Co3--O$j$ and Co4--O$j$ distances with $j=1,\ldots,5$ (see also Fig.~\ref{fig_structure1}), whereas Fig.~\ref{fig_neutrons}~(e) shows the mean $<$Co$i$--O$>$ distances averaged over all oxygen ions. The NPD structural data are overall consistent with those obtained with single crystal XRD. In addition, they show that $<$Co2--O$>$ continuously reduces upon warming in the investigated temperature interval with an accompanying increment of $<$Co4--O$>$, whereas $<$Co1--O$>$ drops by almost $0.02$ \AA\ close to $T=475$~K, nearly matching $<$Co3--O$>$ above this temperature. More specifically, Fig.~\ref{fig_neutrons}(a) indicates anomalies in Co1--O3 and Co1--O2 bond distances at $T=475$ and 495~K, respectively. We also note an anomaly in the Co3--O2 bond distance at $T=495$~K [see Fig.~\ref{fig_neutrons}(c)]. See Fig.~\ref{fig_neutrons_profile} in the SM for the NPD profile at a selected temperature as well as its Rietveld refinement.

\subsection{Raman spectroscopy}

\begin{figure}
	\centering
	\includegraphics{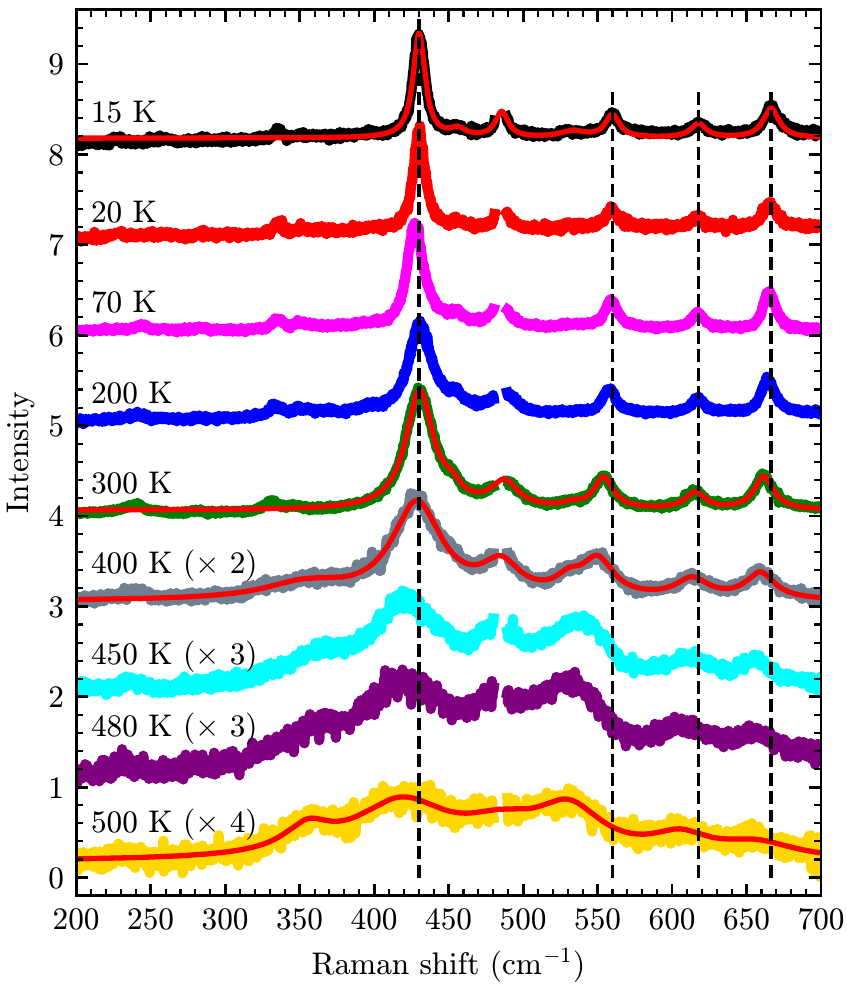}
	\caption{Raman spectra at selected temperatures. The spectra were vertically translated for clarity. The dashed vertical lines mark the frequency positions of selected peaks at 15~K. Fits of the observed spectra with Lorentzian peaks are shown in red solid lines.}
	\label{fig_raman}
\end{figure}

\begin{figure}
	\centering
	\includegraphics{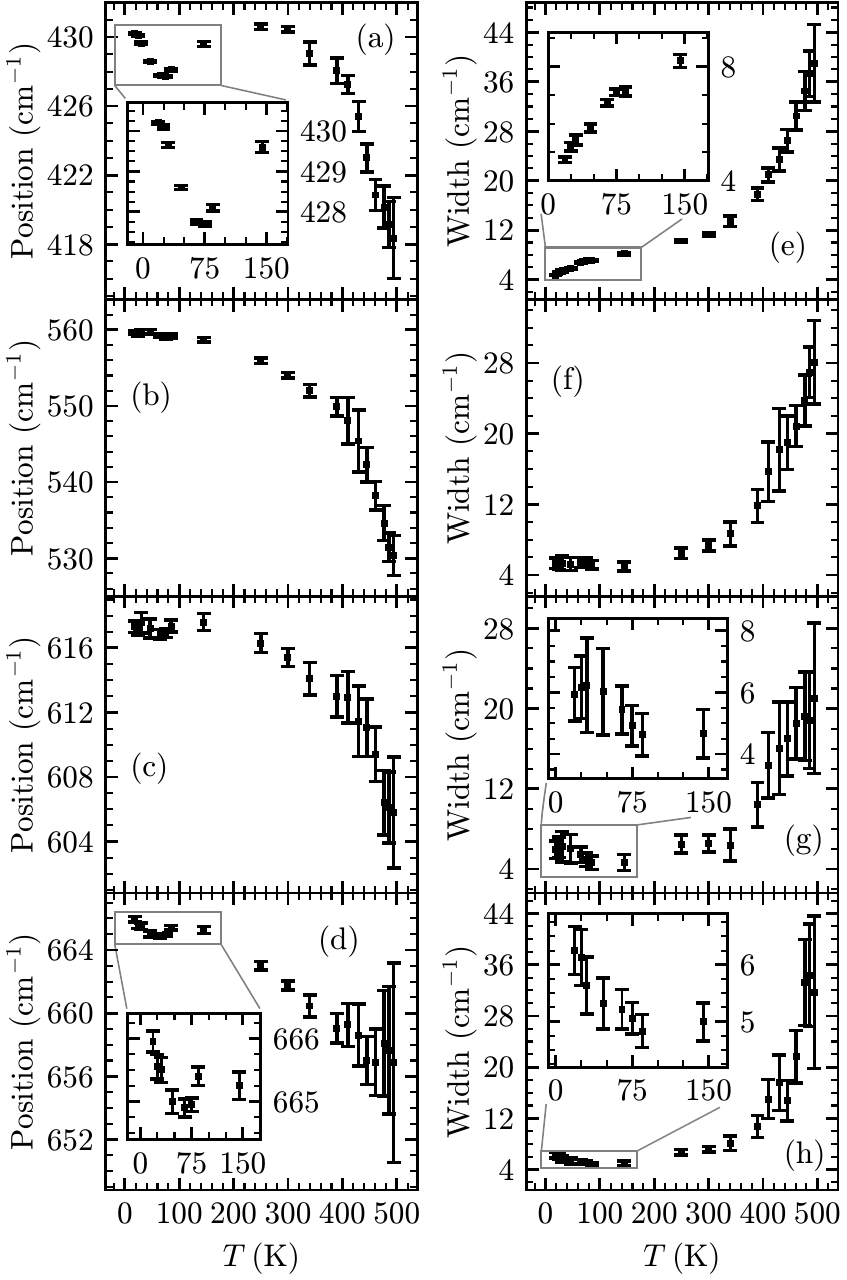}
	\caption{Temperature dependence of (a--d) the frequency positions and (e--h) the corresponding linewidths of the peaks at $\approx 430$, 560, 618 and 665 cm$^{-1}$. The insets show zoom outs that highlight low-temperature anomalies.}
	\label{fig_raman_temp}
\end{figure}

The Raman spectra of single crystalline \chem{\co} are shown in Fig.~\ref{fig_raman} for the spectral region between 200 and 700 cm$^{-1}$ at selected temperatures between 15 and 500~K. At 15~K, sharp peaks are observed at 225, 255, 335, 350, 396, 430, 456, 486, 530, 560, 618, and 665 cm$^{-1}$. At higher temperatures, important changes are observed. The most visible ones are a very large softening ($\approx 30$ cm$^{-1}$) of the 560 cm$^{-1}$ mode and the disappearance of the 335 cm$^{-1}$ mode above $T \approx 300$~K. The strongest peaks were fitted by Lorentzian lineshapes (solid lines in Fig.~\ref{fig_raman}). Figures~\ref{fig_raman_temp}(a--h) show the temperature dependencies of the frequency positions and linewidths of selected modes at 430, 560, 618 and 665 cm$^{-1}$. In common, these modes broaden at a remarkably elevated rate above $\approx 370$~K, reinforcing the picture of an enhanced lattice disorder beyond conventional thermal effects suggested by XRD data (see above). The investigated modes also tend to soften more substantially above this temperature. Beyond these common aspects, we also find some interesting behavior in specific modes. Particularly, the 430~cm$^{-1}$ peak shows an anomalous behavior below $T \approx 200$~K, reaching an frequency minimum at $T = 75$~K [Fig.~\ref{fig_raman_temp}(a), Fig.~\ref{fig_raman_temp}(c), and Fig.~\ref{fig_raman_temp}(d)]. The mode at 430~cm$^{-1}$ sharpens substantially below this temperature [Fig.~\ref{fig_raman_temp}(e)], contrary to the 618 and 665 cm$^{-1}$ modes, which broaden below $T = 75$~K [Fig.~\ref{fig_raman_temp}(g) and Fig.~\ref{fig_raman_temp}(h)].

\subsection{Extended x-ray Absorption Fine Structure (EXAFS)}

\begin{figure}
	\centering
	\includegraphics{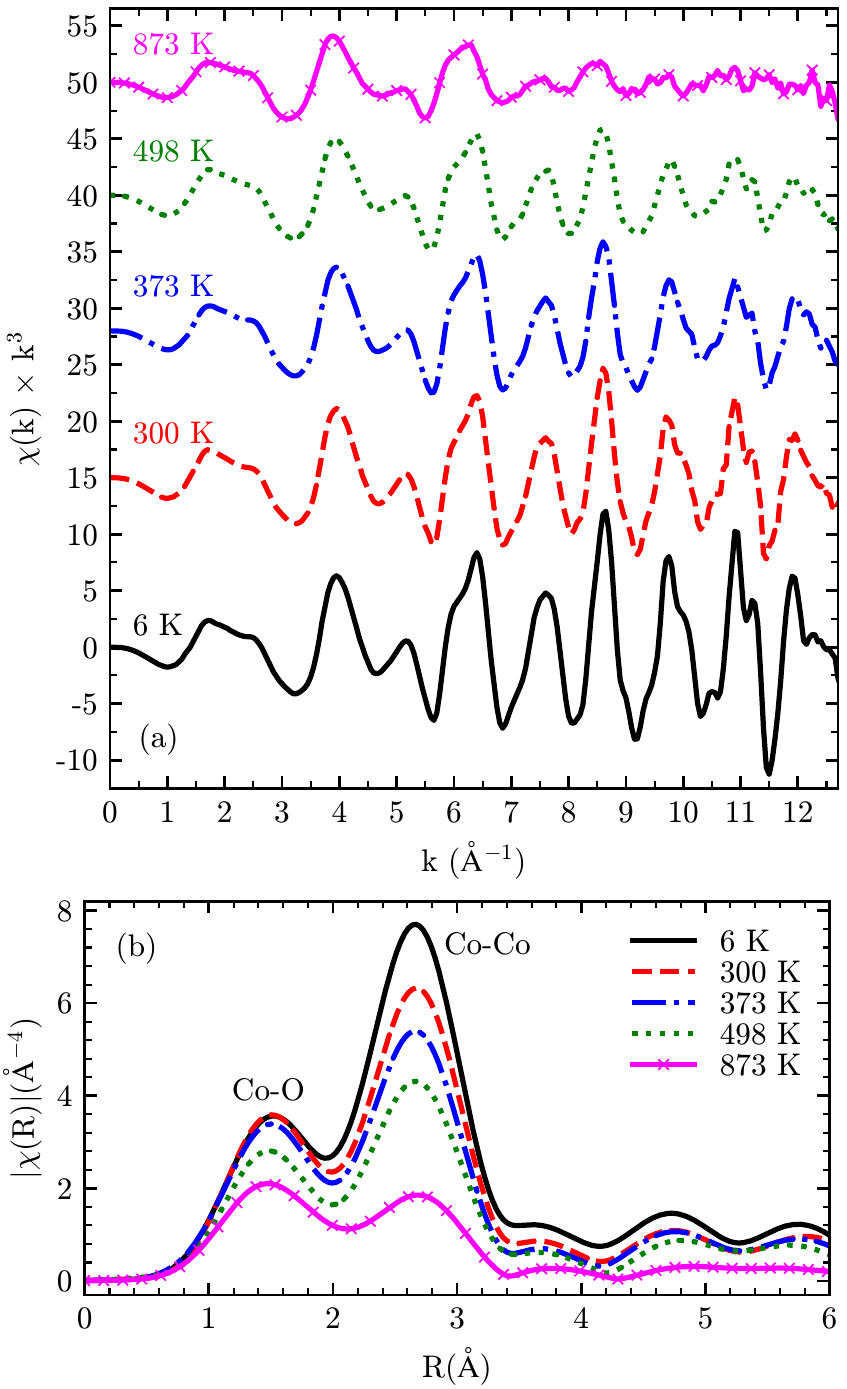}
	\caption{(a) Co $K$-edge $k^{3}$-weighted EXAFS transmission spectra at selected temperatures and (b) their Fourier transform. The spectra were vertically translated in (a) for clarity. The first Co--O and second Co--Co coordination shells are indicated in (b).}
	\label{fig_exafs}
\end{figure}

Figure~\ref{fig_exafs}(a) shows, at selected temperatures, the $k^{3}$-weighted Co $K$-edge EXAFS spectrum of \chem{\co}. Figure \ref{fig_exafs}(b) displays the modulus of the Fourier transformed signal, $| \chi(R) |$, using a gaussian $k$-window $3.6 < k < 10$ \AA$^{-1}$. The first coordination shell, related to Co--O scattering paths, does not change significantly from the base temperature up to $\approx 370$~K [see Fig.~\ref{fig_exafs}(b)], suggesting rigid oxygen octahedra around the Co ions in this temperature range. Above $\approx 370$~K, $|\chi(R)|$ around the first coordination shell starts to dampen, revealing an incremented variance of the Co--O bond lengths with respect to the average value. The second coordination shell, mainly composed by Co--Co scattering paths, shows a different temperature dependence, being continuously damped upon warming.

\begin{figure}
	\centering
	\includegraphics{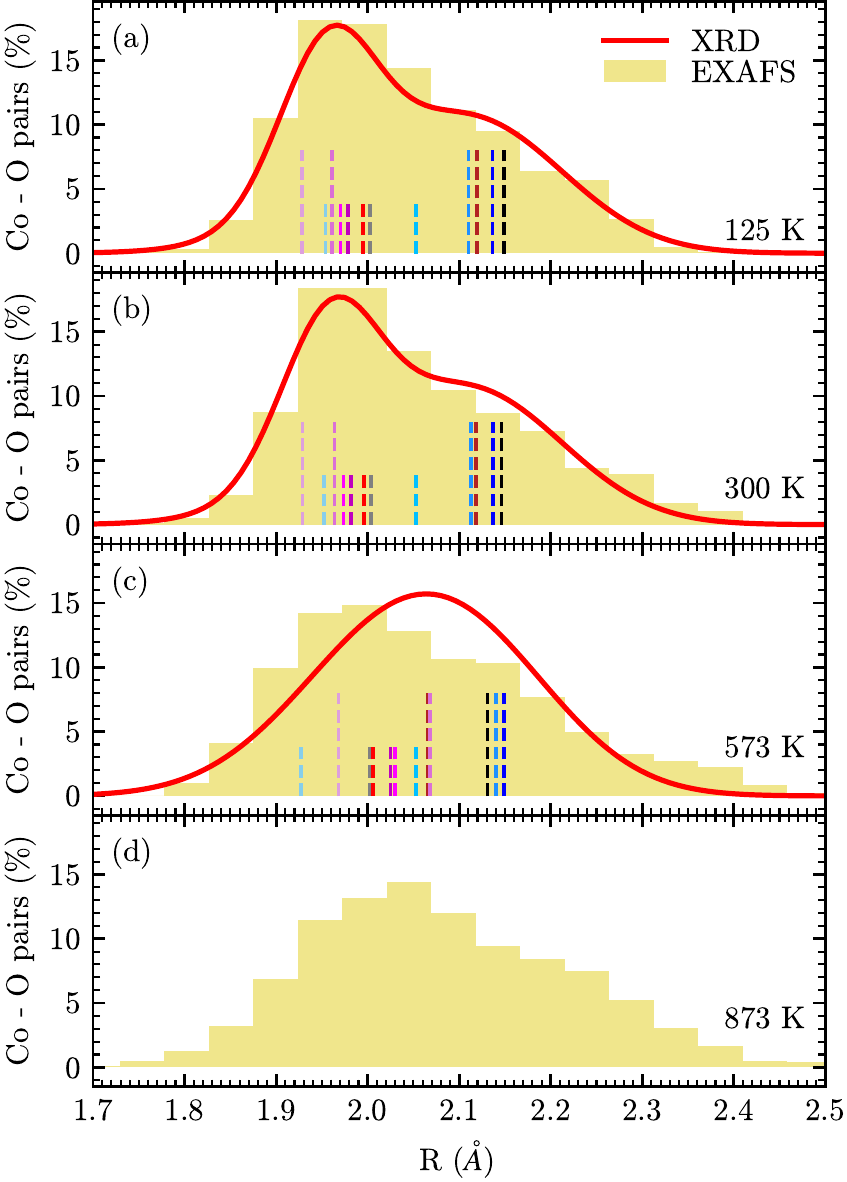}
	\caption{Bars: local Co--O bond length distribution of \chem{\co} obtained from RMC simulation of Co $K$-edge EXAFS data (Fig.~\ref{fig_exafs_fit}) at (a) 125K, (b) 300~K, (c) 573~K, and (d) 873~K. The vertical dashed lines in (a--c) represent the crystallographic Co$i$--O$j$ bond lengths obtained with x-ray diffraction data, where the color scheme is the same as in Figs.~\ref{fig_xrd}(a)--\ref{fig_xrd}(d). The red solid lines in (a--c) are the Gaussian-broadened XRD Co$i$--O$j$ bond length distributions at similar temperatures.}
	\label{fig_rdf}
\end{figure}

The histograms of Figs.~\ref{fig_rdf}(a--d) show the Co--O bond length distribution obtained by the RMC simulated model at 125, 300, 573, and 873~K (refer to Figs.~\ref{fig_exafs_fit}(a) and \ref{fig_exafs_fit}(b) in the SM for the model's calculated EXAFS signal for the respective temperatures). For comparison, the Co--O bond distances inferred from single-crystal XRD data are displayed as vertical dashed lines except at 873~K where XRD data are not available. The solid curves in red are the gaussian broadened XRD bond length distributions. At $T=125$ and 300~K, the Co--O bond distance distributions obtained from EXAFS and XRD data are consistent to each other and strongly asymmetric. At $T=573$~K, the distribution obtained by XRD becomes more symmetric as a consequence of the smaller size differentiation between the Co4O$_6$ and Co2O$_6$ octahedra [see also Fig.~\ref{fig_xrd}(e)]. This substantial evolution with temperature is not followed by the EXAFS Co--O bond length distribution, which remains highly asymmetric and reminiscent of the curves at lower temperatures. At $T=873$~K, this distribution remains asymmetric and similar to the data at $T=573$~K, except for an overall shift towards larger bond lengths possibly due to anharmonicity. The Co--O bond length distribution obtained from the RMC simulation of EXAFS data for several other investigated temperatures from 6 to 873~K are displayed in Fig.~\ref{sup_pdf} in the SM.

\subsection{X-ray Absorption Near Edge Structure (XANES)}

\begin{figure}
	\centering
	\includegraphics{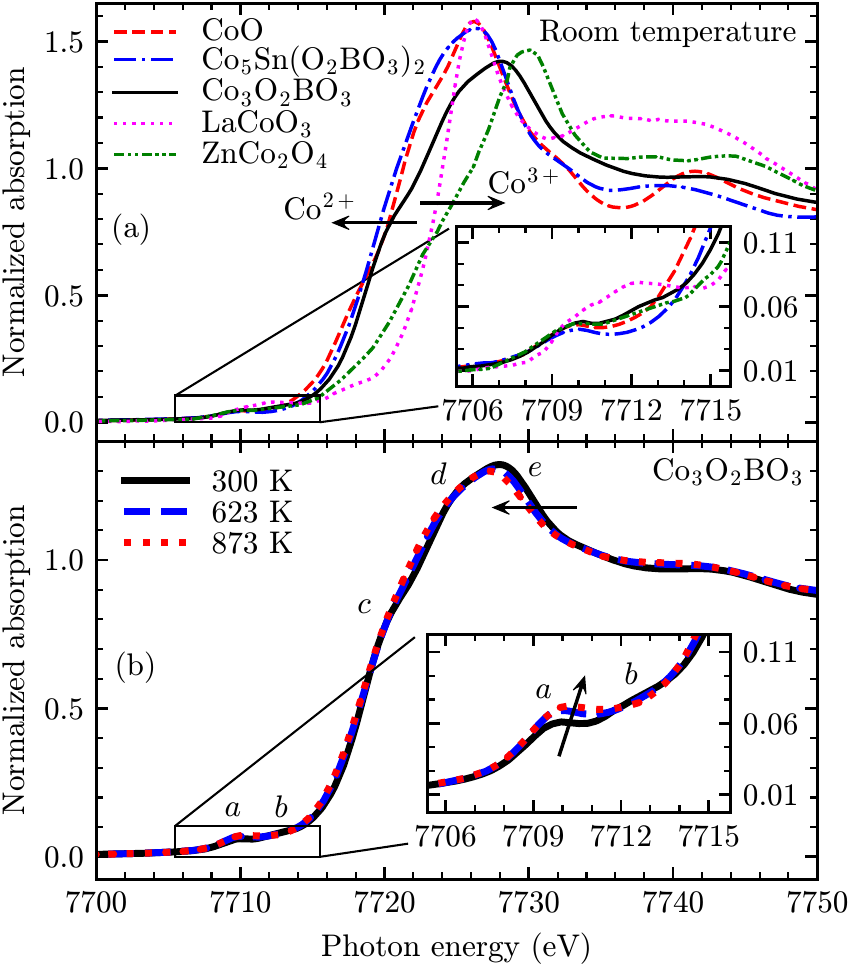}
	\caption{(a) Normalized XAS spectra at the Co $K$-edge of \chem{\co} and reference compounds \chem{\sn}, LaCoO$_3$ and CoO at room temperature measured with membrane samples. Arrows indicate the general notion that higher (or lower) valence states shifts the spectral ``center of gravity'' to higher (or lower) energies. (b) Normalized XAS spectra at the Co $K$-edge of \chem{\co} at 300~K, 623~K and 873~K, where arrows indicate visually detectable changes in the spectra going from low to high temperature.}
	\label{fig_xanes}
\end{figure}

Figure~\ref{fig_xanes}~(a) shows the normalized Co $K$-edge XANES spectra $\mu(E)$ at room temperature of \chem{\co}, reference Co$^{2+}$ compounds \chem{\sn} (also with ludwigite structure \cite{medrano2015nonmagnetic}) and CoO, and reference Co$^{3+}$ compounds LaCoO$_3$ and ZnCo$_2$O$_4$ (where the later was extracted from Ref.~\cite{rong2015zinc}). The most prominent features of \chem{\co} are captured by a simple linear combination of $\frac{2}{3}$ of $\chi(E)$ for \chem{\sn} and $\frac{1}{3}$ of $\chi(E)$ for ZnCo$_2$O$_4$ (see Fig.~\ref{sup_xanes2} in the SM and Ref.~\cite{kazak2021spin}), providing direct experimental confirmation of the $\frac{2}{3} \times$ Co$^{2+}$ + $\frac{1}{3} \times$ Co$^{3+}$ mixed valence inferred from the charge neutrality condition of the chemical formula.

Figure~\ref{fig_xanes}~(b) shows the normalized Co $K$-edge XANES spectra of \chem{\co} at 300~K, 623~K, and 873~K. Spectral features are indicated by letters ($a$--$e$) and arrows indicate the most notable spectral weight changes upon warming. The overall XANES spectral variations of \chem{\co} with temperature are subtle, indicating that the $\frac{2}{3} \times$ Co$^{2+}$ + $\frac{1}{3} \times$ Co$^{3+}$ mixed valence is retained over the entire investigated temperature interval. The temperature dependencies of the energy and spectral weight of selected features, as well as a discussion on their origins, are given in the section~\ref{sec_sup_xanes} of the SM. Such detailed analysis suggests that the charge disorder crossover takes place over a very wide temperature interval, with a tendency for saturation only above $T \sim 800$ K.



\subsection{Differential scanning calorimetry}

Differential scanning calorimetry (DSC) data for our \chem{\co} samples were published earlier, see Ref.~\onlinecite{galdino2019magnetic}. Here, we present DSC data for the other homometallic ludwigite, \chem{Fe_3O_2BO_3}, which will be relevant for our discussion (see below). Figure~\ref{fig_dsc} shows heat flow measurements for a \chem{Fe_3O_2BO_3} single crystal. The most notable feature is the endothermic (exothermic) transformation at 283~K on heating (cooling) associated with the long-range structural/charge-ordering transition, as previously reported \cite{mir2001structural, freitas2008structure}. Remarkably, another endothermic (exothermic) transformation is seen at 500~K in the heating (cooling) curves, indicating an additional high-temperature phase transition for this compound.

\begin{figure}
	\centering
	\includegraphics{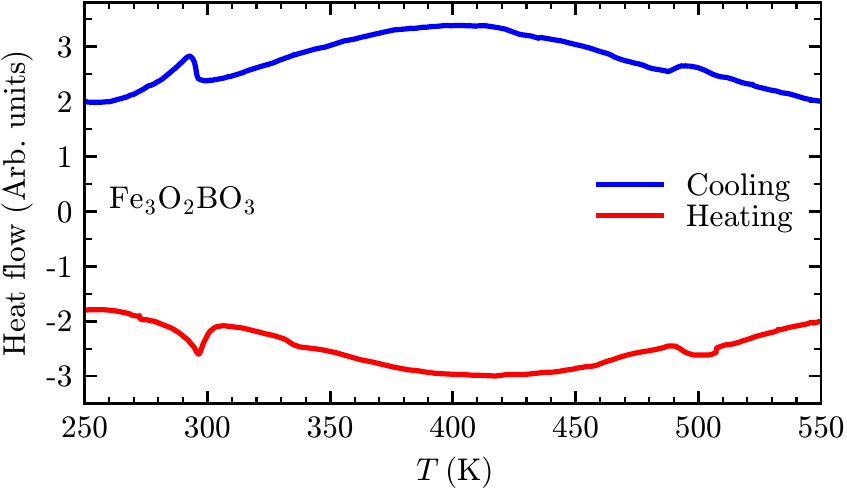}
	\caption{Heat flow as a function of temperature for a \chem{Fe_3O_2BO_3} single crystal, taken on heating and cooling with a rate of 10~K/min.}
	\label{fig_dsc}
\end{figure}

\section{Discussion}

\begin{figure*}
	\centering
	\includegraphics[width=\linewidth]{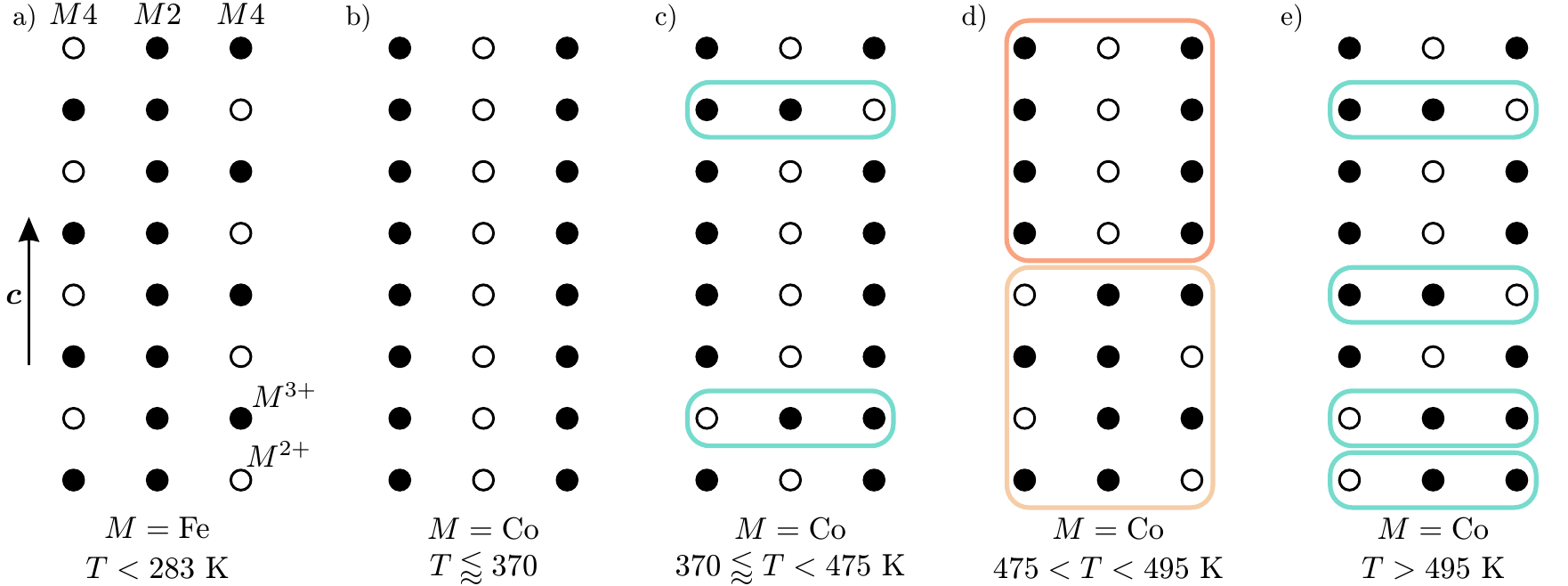}
	\caption{Schematic charge arrangements for $M^{2+}_2M^{3+}$O$_2$BO$_3$ homometallic ludwigites inside the $424$ ladders for $M$=Fe and Co at low temperatures, and for $M$=Co at higher temperatures. Filled and empty symbols represent $M^{3+}$ and $M^{2+}$ ions, respectively. (a) $M$=Fe charge-ordered state ($T<283$~K) \cite{douvalis2002mossbauer, guimaraes1999cation, larrea2001charge, mir2001structural, fernandes2005transport}. (b) $M$=Co charge-ordered state ($T \lesssim 370$~K). (c) $M$=Co charge-ordered state with uncorrelated ``defective rows'' (blue boxes, $370 \lesssim T < 475$~K). (d) The defects become denser upon warming and finally coalesce into defect-rich regions through a phase transition, forming a short-range zig-zag pattern (light red boxes) coexisting with defect-free regions (dark red boxes, $475<T<495$~K). (e) At higher temperatures ($T>495$~K), the thermal energy disrupts the defective-row correlations, returning the system to an uncorrelated defect state that is qualitatively similar to (c) but with higher density of defects, through a reentrant phase transition.}
	\label{fig_chargeorder}
\end{figure*}

In \chem{\co}, the small $<${\it M}4--O$>$ bond length of $\approx 1.95$ \AA\ for temperatures below 370~K, contrasting with 2.07--2.10 \AA\ for $M1$, $M2$, and $M3$ sites [see Fig.~\ref{fig_xrd}(e)], is a clear manifestation of the large degree of charge-order, with the smaller Co$^{3+}$ ions predominantly located at the $M4$ site as previously reported \cite{freitas2008structure,freitas2016magnetism}. It is interesting to compare the charge-ordered configuration along the 424 ladder of $M_3$O$_2$BO$_3$ ludwigite with $M$=Co [see Fig.~\ref{fig_chargeorder}(b)] with the other known homometallic ludwigite ($M$=Fe) at low temperatures [see Fig.~\ref{fig_chargeorder}(a)]. For $M$=Co, the full occupation of the $M4$ site with Co$^{3+}$ ions implies in the formation of infinite Co$^{3+}$ stripes along {\bf c} in the lateral columns of the 424 ladders, whereas, for $M$=Fe, the $M2$ central column is occupied by Fe$^{3+}$ ions and the other Fe$^{3+}$ ions occupy each side of the 424 ladder alternately \cite{mir2001structural}. This leads to a zig-zag charge-order pattern with a corresponding space group change of the crystal lattice to $Pbnm$ and doubling of the {\it c} lattice parameter below $T=283$~K. In common, both charge-ordered patterns have 2$M^{3+}+1M^{2+}$ ions in each and every rung of the 424 ladder. 


The local $M$--O bond length distribution in \chem{\co} obtained with EXAFS is consistent with XRD data at low temperatures [see Figs.~\ref{fig_rdf}(a--b)], with an asymmetric distribution associated to the coexistence of smaller Co$^{3+}$ and larger Co$^{2+}$ ionic radius. At high temperatures, the distance distribution obtained with XRD becomes more symmetrical reflecting the tendency for size equalization of the $M2$ and $M4$ sites. This tendency is not accompanied by the EXAFS local $M$--O distance distribution, which remains largely asymmetric at high temperatures [see Fig.~\ref{fig_rdf}(c)]. These combined data indicate individual Co ions in well defined (Co$^{2+}$ or Co$^{3+}$) oxidation states at all temperatures. At low temperatures, these states are organized in the 424 ladders of the $M$=Co ludwigite structure as $M4^{3+}$--$M2^{2+}$--$M4^{3+}$ with translational invariance along {\bf c}, giving rise to infinite charge stripes [Fig.~\ref{fig_chargeorder}(b)]. Such charge-ordered state is progressively disturbed above $\approx 370$~K, with the extra positive charge at the $M4$ site at low temperatures being, on average, increasingly shared with the $M2$ site upon warming [see Fig.~\ref{fig_chargeorder}(c) and the Co2 and Co4 curves in Fig.~\ref{fig_xrd}(e)]. It should be mentioned that the $M1$ and $M3$ sites do not show a substantial mean size evolution with temperature in comparison to $M2$ and $M4$ [see Fig.~\ref{fig_xrd}(e)], indicating that these sites remain mostly occupied by Co$^{2+}$ over the entire investigated thermal interval.

The charge-order/disorder transformation within the 424 ladders in the Co ludwigite above $T \approx 370$~K is also manifested by a substantial increment of lattice disorder, as revealed by the anomalous behavior of the XRD $\sigma^2$ parameters [see Figs.~\ref{fig_debye}(a--i)] and the phonon linewidths [see Figs.~\ref{fig_raman_temp}(e--h)]. Such enhanced disorder could be indeed anticipated under the assumption that the extra positive charge of Co$^{3+}$ remains localized in the atomic scale at high temperatures, provoking fluctuations in the $M$--O bonds as charges hop from one site to the other. The anomalous enhancement of XRD $\sigma^2$ is more prominent for the O5 oxygen site [see Fig.~\ref{fig_debye}(i)], which is precisely the anion that connects the M2 and M4 cations in the 424 ladder [see Fig.~\ref{fig_structure1}]. As for the Co sites, the lower $\sigma^2$ and the correspondingly higher $\theta_D$ for M4 with respect to the other Co sites below 370~K [see Fig.~\ref{fig_debye}(d)] is likely associated with the smaller $<${\it M}4--O$>$ distance that tends to reduce the Co4 thermal vibration amplitude.

The charge reorganization in \chem{\co} is not accompanied by a structural phase transition as it is observed for other mixed-valence transition metal oxides, e.g., the iron ludwigite at 283~K~\cite{douvalis2002mossbauer, guimaraes1999cation, larrea2001charge, mir2001structural, fernandes2005transport} or the magnetite at 120~K~\cite{verwey1939electronic,Iizumi1982}. Indeed, the ludwigite structure with $Pbam$ space group supports stripe charge-ordered configurations without necessarily reducing the symmetry of the lattice with respect to the disordered configuration, as effectively observed here for \chem{\co}. In this case, the electrostatic potentials associated with the four $M$ sites are different even in the charge-disordered configuration, meaning that the order parameter associated with the average electronic occupation in each site does not become critical and cannot be associated with a phase transition. Instead, a crossover between a low-temperature charge-ordered to a high-temperature disordered configurations might be anticipated, which is consistent with the gradual tendency for average size equalization of Co ions the $M2$ and $M4$ sizes upon warming [see~Figs.~\ref{fig_xrd}(e) and \ref{fig_neutrons}(e)]. However, these considerations do not rule out the possibility of formation of self-organized states in an intermediate length scale and phase transitions involving them. This possibility is actually supported by sharp anomalies in calorimetry and electrical resistivity measurements at 475 and 495~K, as well as anomalies in the $c$ lattice parameter~\cite{galdino2019magnetic}, which appears to be accompanied by subtle bond length anomalies [see Figs.~\ref{fig_neutrons}(a) and (c)]. These transitions must be purely electronic as there is no sign of space group change in \chem{\co} at these temperatures.

A simple picture is able to explain the experimental results at high temperatures. Upon warming above $T \approx 370$~K, thermally-induced defects in the charge-ordered pattern shown in Fig.~\ref{fig_chargeorder}(b) gradually starts to appear, creating `defective rungs' in the 424 ladder such as those displayed in Fig.~\ref{fig_chargeorder}(c), which reduce the average size differentiation between crystallographic sites $M2$ and $M4$ and increases the lattice disorder. Assuming that such defects have mobility along the {\bf c} direction, the energy penalty of each thermally-induced defective rung may be minimized if the defects coalesce into defect-rich nanoregions with a zig-zag charge configuration, similar to the structure found in Fe$_{3}$O$_2$BO$_3$, separated from defect-free regions [Fig.~\ref{fig_chargeorder}(d)]. For a sufficiently large thermal-induced defect density, the energy gain of this segregated phase may overcome its smaller entropy and the segregated-phase may show smaller free energy $F=U-TS$ with respect to the phase with uncorrelated defects [Fig.~\ref{fig_chargeorder}(c)] --- at least within a limited temperature window. Thus, the sequential phase transitions at $T=475$ and 495~K for \chem{\co} are likely transitions from the uncorrelated [Fig.~\ref{fig_chargeorder}(c)] to the segregated [Fig.~\ref{fig_chargeorder}(d)] and back to the uncorrelated defect phases [Fig.~\ref{fig_chargeorder}(e)], i.e., a reentrant transition. The evidently lower charge entropy of the intermediate segregated-phase between $T=475$ and 495~K with respect to the uncorrelated defect phase is supported by differential scanning calorimetry (DSC) data that show an exothermic (endothermic) process upon warming (cooling) in the background-subtracted curve between 475 and 495~K~\cite{galdino2019magnetic}. 
Moreover, the electrical conductivity $\sigma(T)$ of \chem{\co} also supports a reentrant phase transition, as the $\sigma(T)$ curve above $T=495$~K matches the extrapolated behavior of the curve below $T \approx 475$~K (see Fig.~\ref{sup_sigma} in the SM and Ref.~\onlinecite{galdino2019magnetic}).

The above picture also provides insight to the high-temperature behavior of the Fe$_3$O$_2$BO$_3$ homometallic ludwite (see Fig.~\ref{fig_dsc}). Following a previous \textit{ac}-conductivity study \cite{fernandes2005transport}, the configuration above the charge-ordering/structural transition at $T_S=283$~K is most likely formed by regions with short-range ordered zigzag chains. Thus, the feature at $T \approx 500$ K revealed by DSC data may indicate a phase transition to a more entropic state in similarity to the higher temperature transition at $T=495$ K for \chem{\co}. Future diffusive x-ray scattering experiments are necessary to confirm this proposal and obtain more detailed information on the short-range charge arrangements in both Co and Fe homometallic ludwigites. It is worth mentioning that, high-temperature anomalies in DSC data for the iron ludwigite could also be caused by the Fe$^{2+}$ to Fe$^{3+}$ oxidation as suggested for vonsenite and hulsite minerals~\cite{biryukov2020investigation}.

Notice that in the Fe ludwigite, the dimerization of the ladder leads, even in the case of non-interacting electrons, to a Peierls-like insulating phase for 1/3 occupation of the 4-2-4 ladder \cite{latge2002transverse}. On the other hand, in \chem{\co} the proposed charge-ordering in this ladder, with an extra electron per rung, corresponds to a metallic state in the absence of correlations. Then, its insulating behavior requires the presence of electronic correlations. A local Hubbard $U$ suffices to make this material insulating \cite{kazak2021spin} and can provide the high energy scale for the charge-ordering phenomena we observe.

At this point, it is worthwhile to compare our results with those recently reported by Kazak et al. in Ref.~\onlinecite{kazak2021spin}. In such study, mean bond lengths $<${\it M}4--O$> \approx 1.98$~\AA\ and $<${\it M}2--O$> \approx 2.05$~\AA\ are reported at 300~K~\cite{kazak2021spin}. This corresponds to a smaller size differentiation of Co ions located at the $M2$ and $M4$ sites in comparison to our data, suggesting samples with a smaller degree of charge-order at room temperature in comparison to the samples investigated here. This is possibly a consequence of very similar electrostatic potentials at sites $M2$ and $M4$, favoring charge disorder within the 424 ladder that may be triggered even by a small degree of atomic defects such as oxygen or cationic vacancies. In addition, the sharp transitions at 475 and 495~K observed in our samples (Ref.~\onlinecite{galdino2019magnetic}) are not observed on the calorimetry data of Ref.~\onlinecite{kazak2021spin}, which rather show a broad maximum at $T \approx 500$~K. It is possible that even a small degree of lattice defects is sufficient to pin the defective rungs, preventing the formation of the phase segregated state displayed in Fig.~\ref{fig_chargeorder}(d). Finally, one might attempt to ascribe the increment of $<${\it M}4--O$>$ distance upon warming to a Co$^{3+}$ spin state crossover from the low-spin to intermediate- or high-spin configurations. Nonetheless, this scenario is not able to explain the simultaneous decrease of $<${\it M}2--O$>$ distance upon warming. Moreover, previous magnetic susceptibility measurements showed that \chem{\co} follows a Curie-Weiss behavior above room temperature~\cite{galdino2019magnetic}, indicating a stable spin configuration for the Co ions in this temperature range.

In addition to the enhanced lattice disorder above $T \approx 370$~K, Raman spectroscopy indicates intermediate-temperature phonon anomalies at $T \sim 75$ K, see Figs.~\ref{fig_raman_temp}(a--h)]. The 430~cm$^{-1}$ mode also show a frequency anomaly below 200~K [see Fig.~\ref{fig_raman_temp}(a)]. These results indicate that additional physics not directly related to the partial melting of the Co charge-ordering above $\sim 370$~K (see above) or to the long range magnetic ordering below $T_C=42$~K~\cite{freitas2008structure,freitas2016magnetism} is at play at intermediate temperatures. Possibilities are the presence of short-range magnetic correlations that may affect the phonon energies through the spin-phonon coupling effect~\cite{Granado1999}, and/or a Co$^{3+}$ spin state instability that may also affect the lattice dynamics. These effects at the intermediate temperature range between $T_C$ and room temperature deserve further examination.

\section{Conclusions}

In summary, we investigated the crystal, local atomic, and local electronic structures, as well as the vibrational properties of \chem{\co} from 6 to 873~K. XANES and EXAFS indicate a lattice of Co ions with well defined Co$^{2+}$ and Co$^{3+}$ oxidation states at all temperatures. The Co--O bond lengths obtained by XRD and NPD indicates that Co$^{3+}$ ions occupy the $M4$ site at low temperatures, whereas an increasing charge disorder takes over above $\approx 370$~K, leading to an abnormal level of lattice disorder induced by local defects in the charge-ordered pattern. A phase-segregated state with short-range correlated rungs within the 424 ladders, where zig-zag pattern regions are intercalated with charge stripe domains, is proposed to explain the sharp phase transitions previously reported at 475 and 495~K in this compound~\cite{galdino2019magnetic}. Above this temperature, a charge-defect-rich reentrant phase with uncorrelated defects takes place. A phase transition at $T \approx 500$~K is also observed by DSC data in the other known homometallic ludwidge \chem{Fe_3O_2BO_3}, which may indicate that phase transitions involving short-range charge-ordered states are not restricted to \chem{\co}. Finally, Raman spectroscopy also uncovers additional phonon anomalies at $T \sim 200$ K and $T \sim 75$ K in \chem{\co} presumably related to magnetic correlations or Co$^{3+}$ spin crossover that calls for further examination.

\section{Acknowledgments}
This research used resources of the Brazilian Synchrotron Light Laboratory (LNLS), an open national facility operated by the Brazilian Centre for Research in Energy and Materials (CNPEM) for the Brazilian Ministry for Science, Technology, Innovations and Communications (MCTIC). The XAFS2 beamline staff is acknowledged for the assistance during the experiments. We also acknowledge Institut Laue-Langevin for granting beam time under Proposal No.~5-24-630. D.C. Freitas and M.A. Continentino are grateful for the support from CNPq and FAPERJ agencies. C.B. Pinheiro thanks FAPEMIG. This work was also funded by FAPESP Grants 2018/20142-8 and CNPq Grants 134752/2016-3 and 142555/2018-5, Brazil.

\bibliography{main}

\cleardoublepage

\onecolumngrid

\section*{Supplemental material for:}

\begin{center}
\large\textbf{Structural and spectroscopic investigation of the charge-ordered, short-range ordered and disordered phases of the \ensuremath{\boldsymbol{\mathrm{Co_3O_2BO_3}}} ludwigite}
\end{center}bibtex main.kazak2011superexchange





\renewcommand\thefigure{S\arabic{figure}}
\setcounter{figure}{0}

\renewcommand\thetable{S\arabic{table}}
\setcounter{table}{0}

\renewcommand\thesection{S-\Roman{section}}
\setcounter{section}{0}

\section{Single crystal x-ray diffraction}

\begin{figure}[h!]
	\centering
	\includegraphics{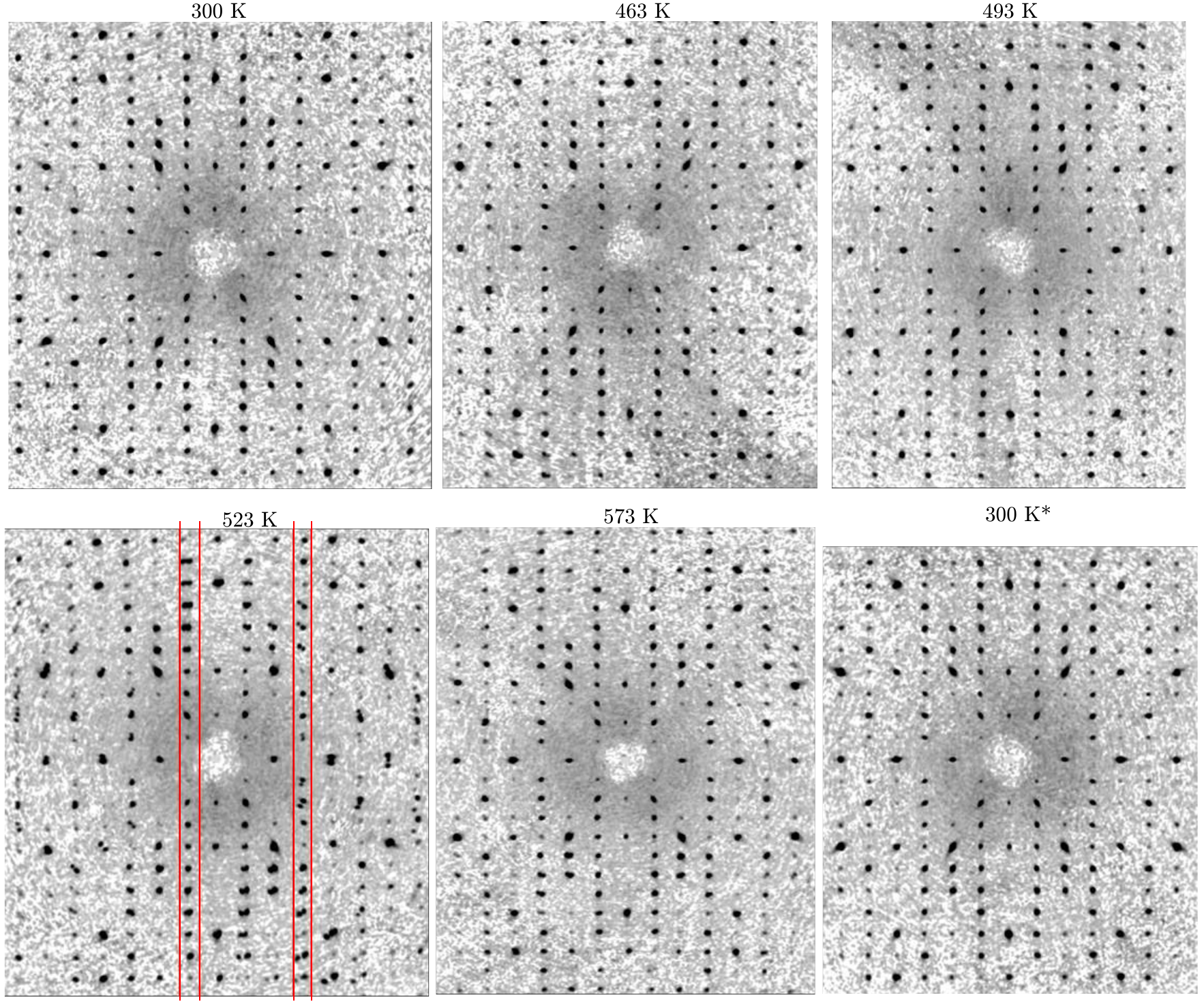}
	\caption{Reconstructed reciprocal space (HK0) layers of the \chem{\co}, where the vertical axis corresponds to $b$* and the horizontal axis to $a$*. At 523~K, the reconstructed layer presents reflections splitting suggesting a twining. Nevertheless, data could be indexed, integrated, and refined properly. Data was collected upon heating from 300~K to 573~K, where the reconstructed layer at 300~K* was obtained after cooling down the sample. Red lines are just guides to the eye. 
	}
	\label{sup_twinning}
\end{figure}

\begin{table}[h!]
\centering
\begin{tabular}{ ||p{3.3cm} |p{2.7cm}| p{2.7cm}|  p{2.7cm}| p{2.7cm}| p{2.7cm}|| }
\hline
\multicolumn{6}{||l||}{Crystal data}\\
\hline
Temperature & 130 K & 300 K & 523 K* & 573 K* & 300 K (cooling)* \\ 
\hline
$a$, $b$, $c$ (\AA)& 9.3123(7) & 9.3031(5) & 9.234(1) & 9.259(1) & 9.3043(7)\\                     & 11.9372(9) & 11.9485(6)&12.009(1) &12.072(1) &11.9509(9)\\
                  & 2.9606(2) & 2.9638(1) & 3.0064(3) & 3.0231(3) & 2.9646(2)\\
\hline
V (\AA$^3$) & 329.11(4) & 329.36(3) & 333.36(6) & 337.88(6) & 329.65(4) \\
\hline
No. of reflections for cell measurement & 5240 & 4003 & 1480 & 1483 & 1491 \\
\hline
$\theta$ range ($^\circ$) for cell measurement & 2.8--33.2 & 2.8--33.1 & 2.8--29.3 & 2.7--29.0 & 2.7--29.1 \\
\hline
$\mu$ (mm$^{-1}$) & 14.83 & 14.81 & 14.64 & 14.44 & 14.80 \\
\hline
Crystal size (mm) & 0.08$\times$0.05$\times$0.03& 0.08$\times$0.05$\times$0.03& 0.48$\times$0.07$\times$0.05& 0.48$\times$0.07$\times$0.05& 0.48$\times$0.07$\times$0.05\\
\hline
\multicolumn{6}{||l||}{}\\
\hline
\multicolumn{6}{||l||}{Data Collection}\\
\hline
Diffractometer& Bruker APEX-II & Bruker APEX-II & Xcalibur, Atlas, & Xcalibur, Atlas,& Xcalibur, Atlas,\\
                &   CCD         & CCD & Gemini ultra& Gemini ultra&Gemini ultra\\
\hline
Absorption correction&Multi-scan &Multi-scan &Analytical&Analytical&Analytical\\
                & SADABS& SADABS&CrysAlis PRO&CrysAlis PRO&CrysAlis PRO\\
                 &       &       & 1.171.38.46& 1.171.38.46&1.171.38.46\\
\hline
$T_{\mathrm{min}}$, $T_{\mathrm{max}}$& 0.364, 0.495&0.393, 0.497&0.149, 0.547&0.129, 0.547&0.153, 0.546\\
\hline
No. of measured, independent and observed [$I > 2\sigma(I)$] reflections&21973, 742, 624&6420, 838, 770&5384, 533, 444&5427, 533, 444&5290, 513, 432\\
\hline
$R_\mathrm{int}$&0.086&0.031&0.078&0.082&0.063\\
\hline
$\theta$ values ($^\circ$)&2.8--33.3&2.8--35.0&2.8--29.6&2.8--29.5&2.8--29.4\\
\hline
$(\sin\theta/\lambda)_\mathrm{max}$ (\AA$^{-1}$)&0.773&0.806&0.695&0.692&0.690\\
\hline
\multicolumn{6}{||l||}{}\\
\hline
\multicolumn{6}{||l||}{Refinement}\\
\hline
${R[F^2>2\sigma(F^2)]}$, $wR(F^2)$, $S$&0.023, 0.046,1.07&0.015, 0.033, 1.13&0.030, 0.070,1.17&0.033, 0.074, 1.19&0.028, 0.064, 1.22\\
\hline
No. of reflections&742&838&533&533&513\\
\hline
No. of parameters&58&58&58&55&54\\
\hline
$\Delta\rho_\mathrm{max}$, $\Delta\rho_\mathrm{min}$ (e \AA$^{-3}$)&0.79, -0.86&0.71, -0.74&1.34, -1.39&1.27, -1.42&0.78, -1.19\\
\hline
\end{tabular}
\caption{Structure refinement parameters for \chem{\co} samples at selected temperatures. Experiments were carried out with Mo K$\alpha$ radiation ($\lambda=0.717073$ \AA). Refinements were performed using an orthorhombic unit cell and space group $Pbam$ with $Z=4$ and $Mr=267.60$. Measurements performed on a different single crystal are marked with an asterisk (*). The complete data can be accessed through Cambridge Crystallographic Data Center (CCDC) deposition numbers 2094101--2094115.}
\label{sup_table}
\end{table}

\clearpage
\section{Neutron powder diffraction}

\begin{figure*}[h]
	\centering
	\includegraphics{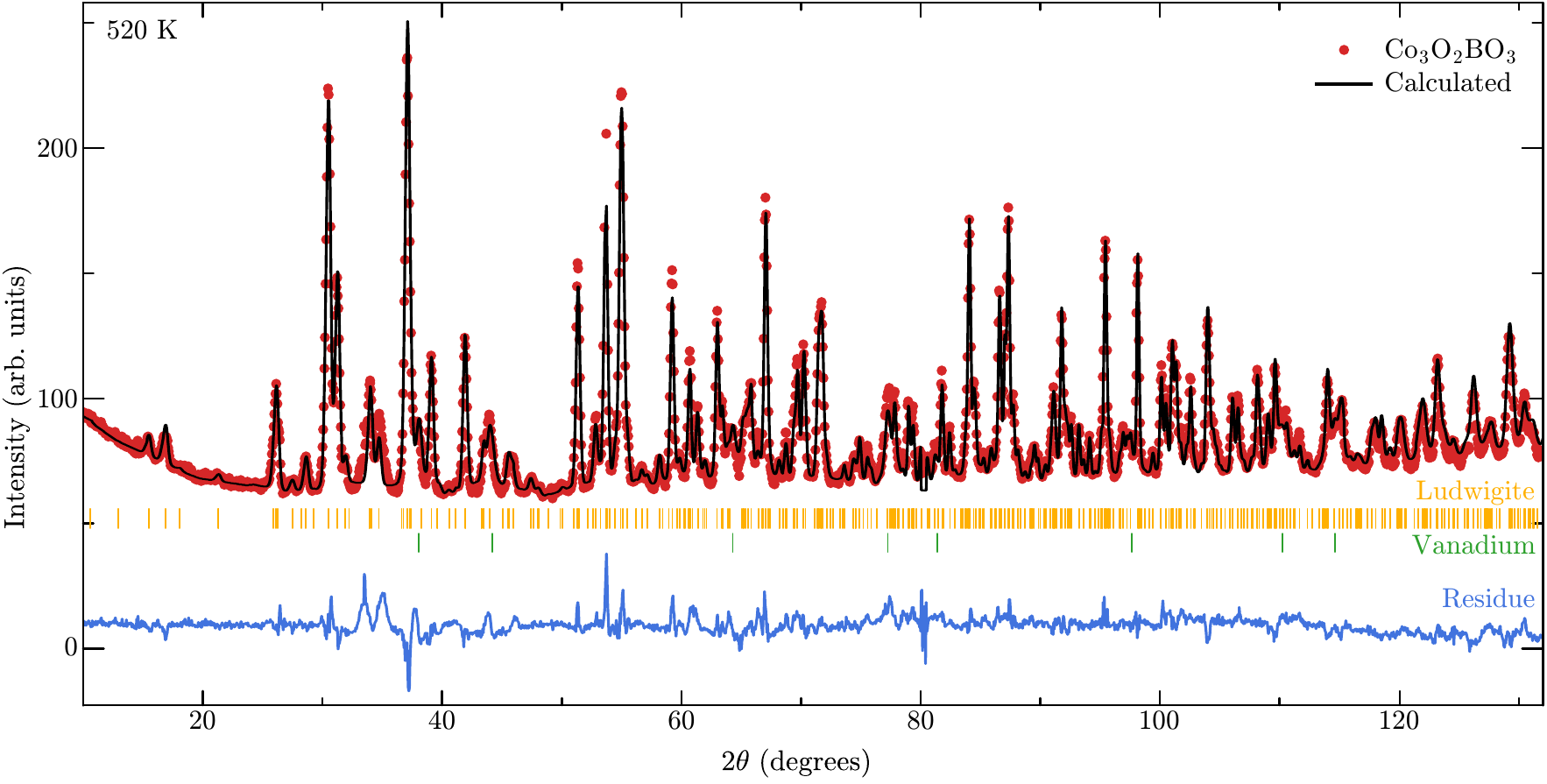}
	\caption{Neutron powder diffraction data of \chem{\co} at $T = 520$~K with $\lambda=1.36$ \AA (red circles) along with a Rietveld refinement (black line). The short vertical bars correspond to the Bragg positions of the main ludwigite phase (space group $Pbam$) and of the vanadium can. The difference curve is shown as a solid blue line at the bottom.}
	\label{fig_neutrons_profile}
\end{figure*}

\clearpage
\section{Extended x-ray Absorption Fine Structure (EXAFS)}

\begin{figure}[h]
	\centering
	\includegraphics{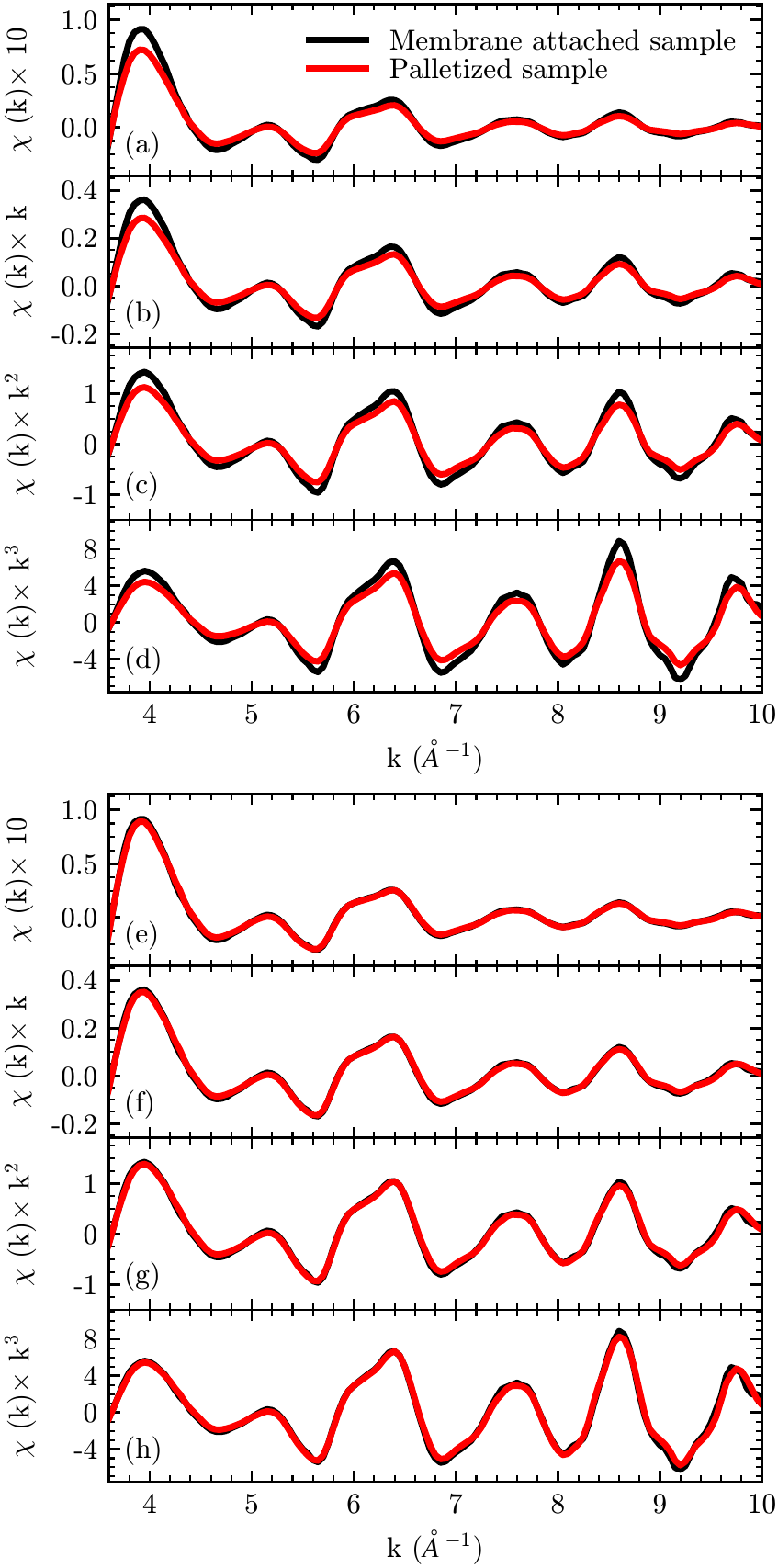}
	\caption{Co $K$-edge EXAFS transmission spectra and its $k$-, $k^{2}$, and $k^{3}$-weighted signal of \chem{\co} at 300~K for membrane attached samples and palletized samples (a--d) before and (e-h) after correction by a 1.23 multiplicative factor for the measurements on the pellet sample to correct for x-ray leakage effects \cite{goulon1982experimental}.}
	\label{sup_exafs}
\end{figure}

\begin{figure}
	\centering
	\includegraphics{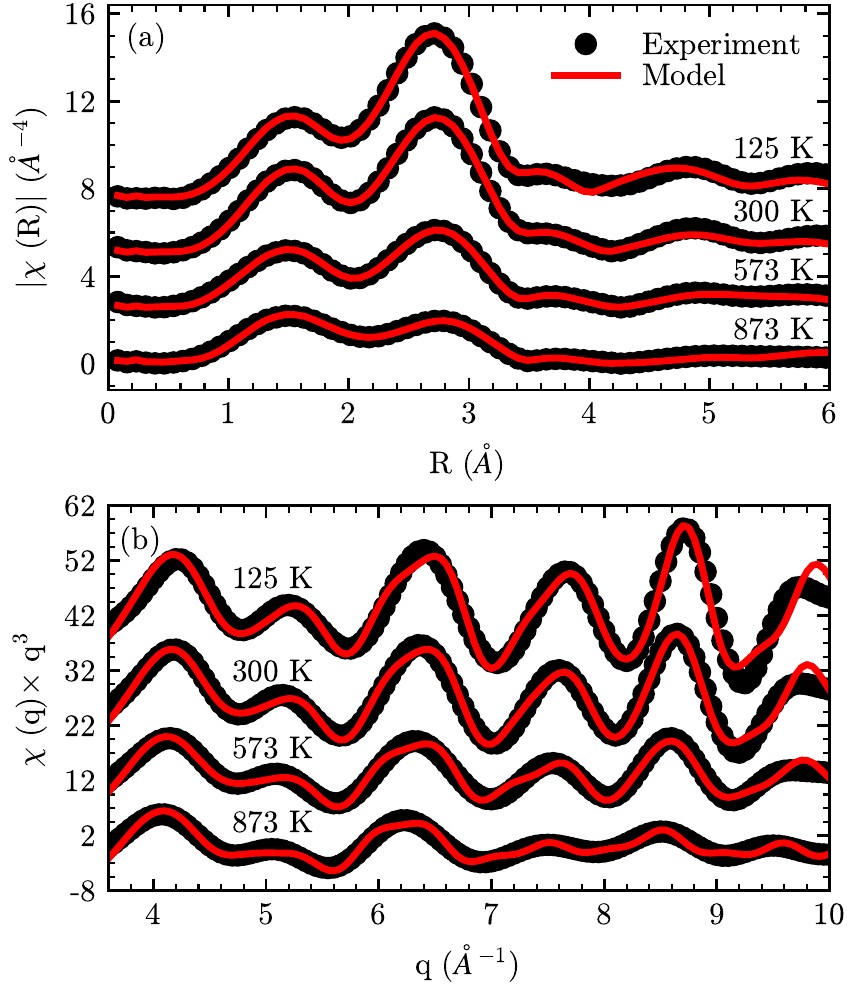}
	\caption{(a) Co $K$-edge $k^{3}$-weighted Fourier-transformed EXAFS spectra at selected temperatures and (b) their back Fourier transform spectra using a gaussian $R$-window $1 < R < 4.5$ \AA\ at selected temperatures. Symbols represent experimental data and solid lines are the results of RMC simulation. Spectra were vertically translated for clarity. The simulated curves reproduces reasonably well the observed features within the designated window in $R-$space and below $\approx 9$ \AA$^{-1}$ in $q-$space. }
	\label{fig_exafs_fit}
\end{figure}

\begin{figure}[H]
	\centering
	\includegraphics{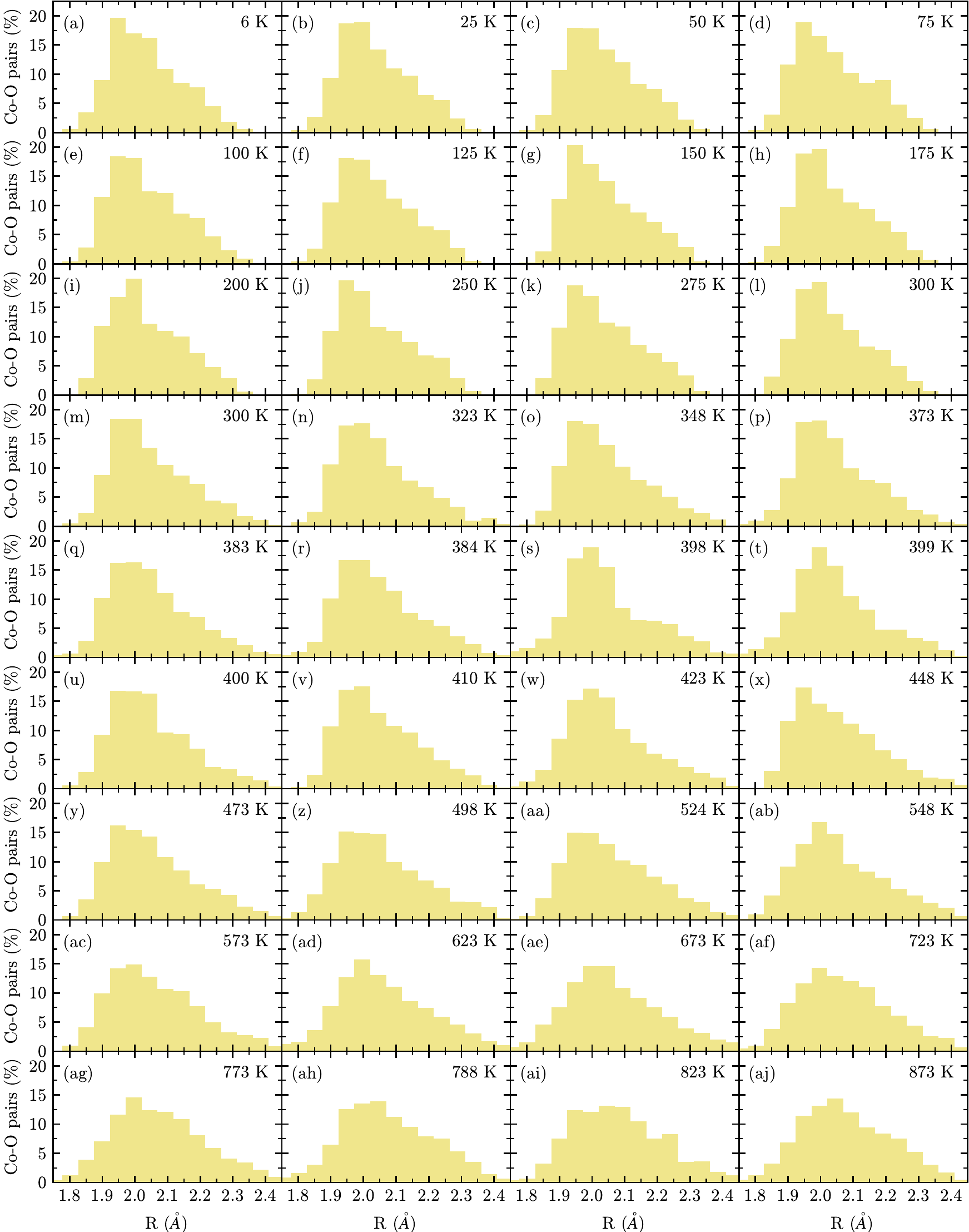}
	\caption{Local Co–O bond length distribution of \chem{\co} at temperatures from 6 to 873~K obtained from RMC simulation of Co $K$-edge EXAFS data.}
	\label{sup_pdf}
\end{figure}

\section{Electrical transport (electrical conductivity)}

\begin{figure}[h!]
	\centering
	\includegraphics{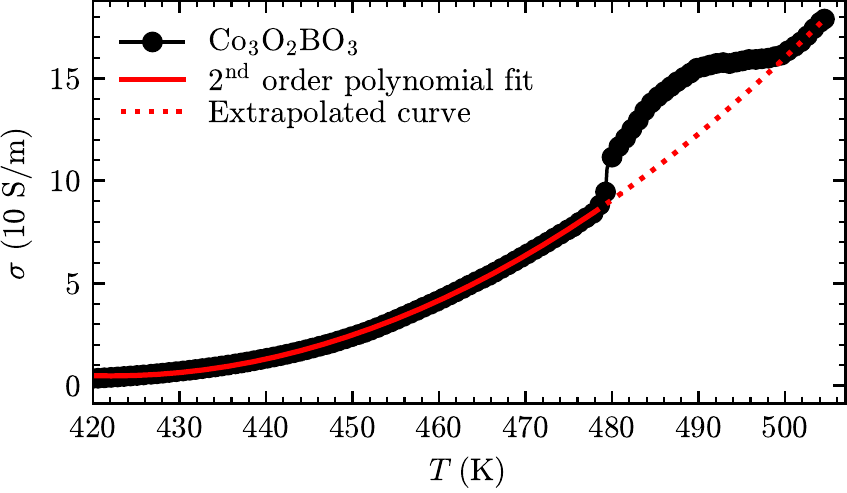}
	\caption{Electrical conductivity $\sigma$(T) of \chem{\co} from Ref.~\onlinecite{galdino2019magnetic}, measured along with the \textbf{c} direction on heating. The solid red line is a 2$^{\mathrm{nd}}$ order polynomial fit from 420 to 478~K, where the curve was extrapolated for higher temperature values (dotted curve).}
	\label{sup_sigma}
\end{figure}

\section{X-ray Absorption Near Edge Structure (XANES)}
\label{sec_sup_xanes}

The XANES spectrum of \chem{\co} has a rich structure with two distinguishable features in the pre-edge region as can be seen in Figs.~\ref{fig_xanes}(a) and \ref{fig_xanes}(b) in the main text. Particularly, the two pre-edge features $a$ and $b$ have been extensively studied by de Groot and Vank\'{o} et al. using RIXS and HERFD-XANES in other cobalt oxides~\cite{vanko2008intersite, de20091s}. Feature $a$ is attributable to a $1s-3d$ quadrupolar transition, whereas feature $b$ has been ascribed to a non-local dipolar excitation related to the $M(4p)$O$(2p)M’(3d)$ intersite hybridization between two neighboring Co ions. This non-local excitation is described as a dipole $1s-4p$ transition to the $4p$ character of the $3d$-band and is related to $M(4p)O(2p)M’(3d)$ intersite hybridization between two metal Co ions in a octahedral environment connected by an oxygen, being only observed for Co compounds with sufficiently short Co--O bond length. In \chem{\co}, we may infer that this intersite hybridization is related with the short Co--O bond length of the Co$^{3+}$ ions at the Co4 site since the Co$^{2+}$ reference compound \chem{\sn}~\cite{medrano2015nonmagnetic, galdino2019magnetic} do not present feature $b$ [see Fig.~\ref{fig_xanes}(a) in the main text]. In the Co$^{3+}$ reference compound \chem{LaCoO_3}, this feature is much more intense when compared to \chem{\co} because the former has corner-sharing octahedra that facilitates hybridization of neighboring Co ions with the same O orbital~\cite{vanko2008intersite, de20091s}.

\begin{figure}[h!]
	\centering
	\includegraphics{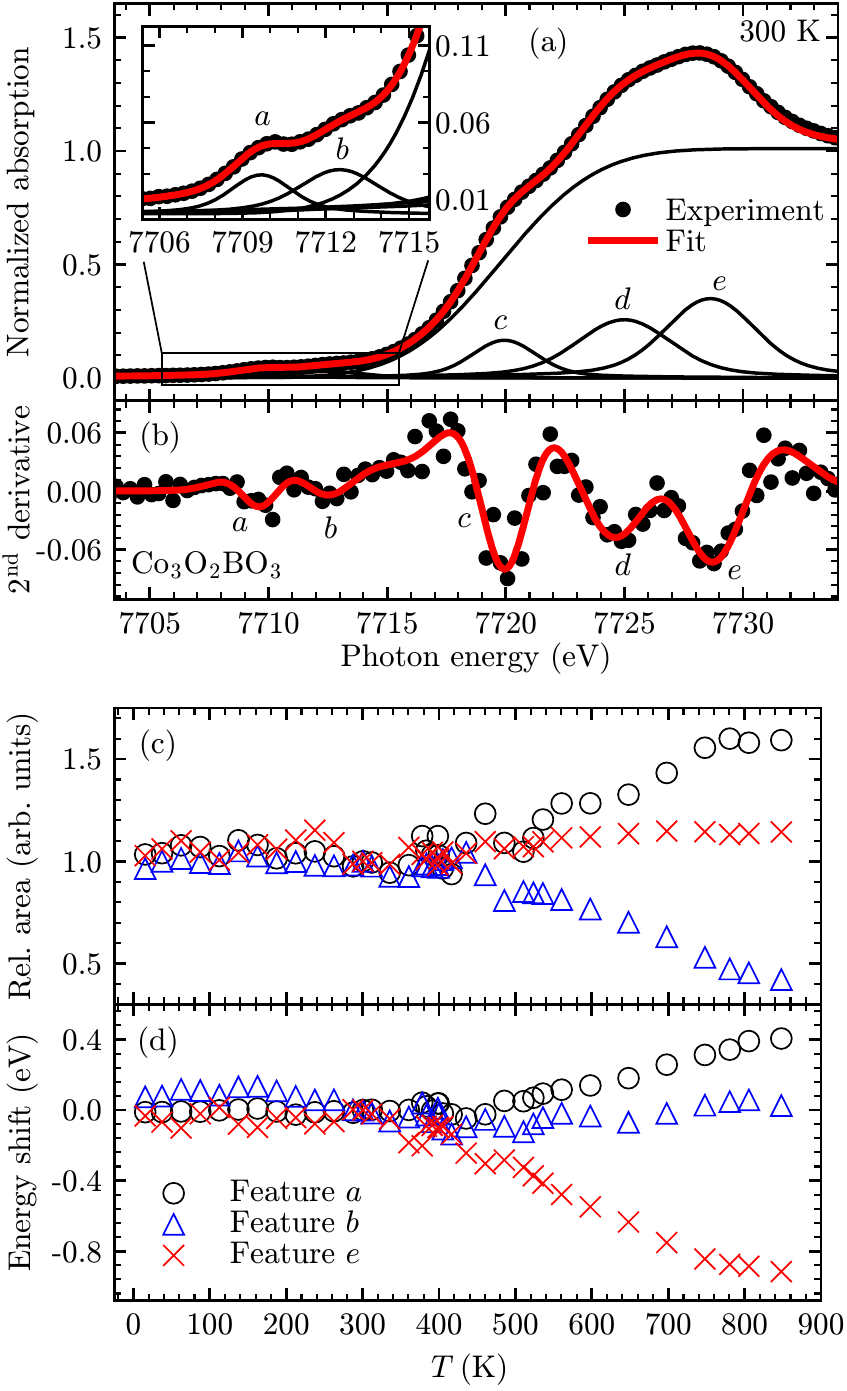}
	\caption{(a) Fit of the XANES spectrum \chem{\co} and (b) its second derivative at 300~K. The inset shows a zoom of the pre-edge region. The spectral features included in the fit are indicated by letters $a-e$. (c,d) Temperature dependence of the (c) integrated intensities and (d) energy shifts of features $a$, $b$ and $e$ relative to the corresponding values at room temperature.}
	\label{sup_xanes}
\end{figure}

We performed a fit of XANES spectra up to $\approx$ 15 eV above the edge [see Fig.~\ref{sup_xanes}(a)]. Five peak-like features are found with the help of the second derivative spectrum [see Fig.~\ref{sup_xanes}(b)]. Feature $c$ is the hardest to visualize as its center roughly matches the center of the arctangent step function. Still, spectral details close to $E \approx 7717$ eV cannot be appropriately fitted without including this feature.
The fitted areas of the selected XANES spectral features $a$, $b$ and $e$ relative to the their values at room temperature are shown in Fig.~\ref{sup_xanes}(c). Figure~\ref{sup_xanes}(d) shows the corresponding energy shifts. These features and the XANES spectrum as a whole remain stable from low temperatures up to $T \approx 400$~K. Above this temperature, the non-local pre-peak $b$ fades away in opposition to the local pre-peak $a$. Also, peak $e$ show a gradual energy shift of -0.9 eV up to $\approx 800$~K, whereas pre-peak $a$ shows a energy shift of $\approx 0.4$ eV. Finding a detailed explanation for each of these trends with temperature is beyond the scope of this work, however it is reasonable to associate the XANES spectral changes above room temperature to the gradual charge-order/disorder crossover in the 424 ladder discussed in the main text. Thus, the temperature dependence of the XANES data at higher temperatures indicates that the charge-order/disorder crossover takes place over a very wide temperature interval, with a tendency for saturation only above $T \approx 800$~K [see Figs.~\ref{sup_xanes}(c) and \ref{sup_xanes}(d)]. No anomaly associated with the sharp transitions at $T=475$ and $495$ K \cite{galdino2019magnetic} is noticed in our XANES spectra.

\begin{figure}[h!]
	\centering
	\includegraphics{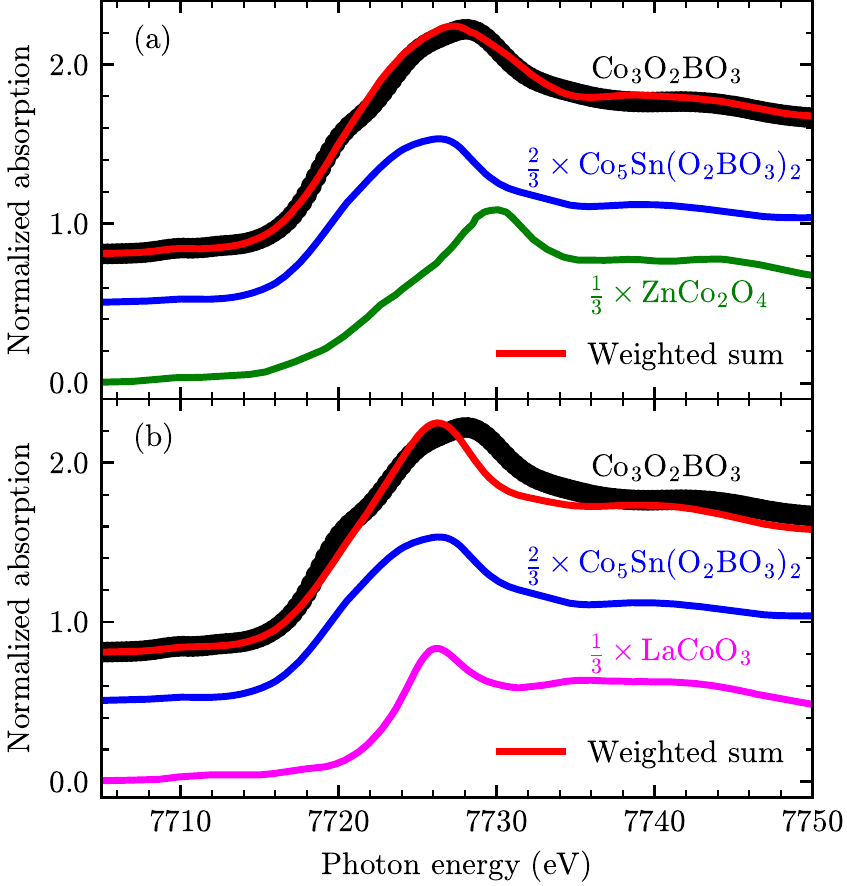}
	\caption{XANES spectrum of \chem{\co} compared with the weighted sum $\frac{2}{3}$ of the Co$^{2+}$ reference \chem{\sn} spectrum plus $\frac{1}{3}$ of the spectrum of the Co$^{3+}$ references (a) ZnCo$_2$O$_4$ and (b) LaCoO$_3$.}
	\label{sup_xanes2}
\end{figure}

\end{document}